\documentclass[%
reprint,
amsmath,
amssymb,
aps,
pra,
]{revtex4-2}

\usepackage{graphicx}
\usepackage{dcolumn}
\usepackage{bm}
\usepackage{units}

\begin{document}


\title{Unambiguous measurement in an unshielded microscale magnetometer with sensitivity below 1 pT/rHz}%

\author{Hamish A. M.~Taylor, Christopher C.~Bounds, Alex Tritt, and L. D.~Turner}
\affiliation{%
 School of Physics \& Astronomy\\
 Monash University, Victoria 3800, Australia
}%

\date{\today}

\begin{abstract}
\noindent 
Cold atom magnetometers exploit a dense ensemble of quanta with long coherence times to realise leading sensitivity on the micrometer scale. Configured as a Ramsey interferometer, a cold atom sensor can approach atom shot-noise limited precision but suffers from fringe ambiguity, producing gross errors when the field falls outside a narrow predefined range. We describe how Hilbert-demodulated optical magnetometry can be realised on cold atom sensors to provide field measurements  both precise and unambiguous. Continuous reconstruction of the Larmor phase allows us to determine the dc magnetic field unambiguously in an unshielded environment, as well as measure ac variation of the field, in a single shot. The ac measurement  allows us to characterize, and then neutralise, line-synchronous magnetic interference, extending reconstruction times. Using $1.6 \times 10^6$ $^{87}$Rb atoms in a volume of $(68 \,\mathrm{\mu m})^3$, we measure a test field to be $\unit[86.0121261(4)]{\mu T}$ in a single shot, achieving dc sensitivity of $\unit[380]{fT}$ in a duration of $\unit[1000]{ms}$. Our results demonstrate that Hilbert-demodulated optical readout yields metrologically-significant sensitivity without the fringe ambiguity inherent to Ramsey interferometry.
\end{abstract}
       
\maketitle

\section{Introduction}

Reconstructing magnetic fields at high spatial resolution and sensitivity underpins many modern applications, including medical diagnostics and imaging~\cite{Belfi,Xu}, and material analysis~\cite{wilson2}. The long coherence time and small size of ultracold atomic and BEC magnetometers makes them strong candidates for sensitive dc magnetometry~\cite{vengalattore,Behbood,Yang} and, increasingly, ac and rf magnetometry~\cite{eto,cohen} on the microscale. While not setting records in terms of absolute sensitivity, the volume-normalised sensitivity of these sensors approaches the proposed $\hbar$ ``limit'' ~\cite{mitchell}, making them highly useful in applications where sub-mm spatial resolution with sufficient sensitivity is key to making precise and accurate measurements of physical and biological systems on the microscale. It is in this domain that we find new high-value applications, such as semiconductor fault analysis~\cite{Chang}, and the detection of currents during the transition of high-$T_c$ superconductors~\cite{Berg}. Conventionally, these magnetometers conclude with a projective measurement of the spin populations after the sensing period, from which the spin dynamics during the sensing period can be inferred. By contrast, thermal atomic vapor magnetometers~\cite{griffith,kominis,sheng,dang,bloom,chalupczak} employ continuous~\cite{Budker} or pulsed~\cite{Borna} optical readout of the Larmor frequency of the atomic cloud in order to impute a magnetic field magnitude.

In the case of Ramsey measurement, arguably the simplest sensing protocol, the atomic spin begins in a spin down eigenstate before being tipped transverse to the field by a resonant rf pulse produced by a local oscillator, beginning Larmor precession. After some sensing time, a second rf pulse converts the accumulated Ramsey phase --- the difference between the Larmor phase and the local oscillator phase --- to a population difference of the spin states. Measuring these populations allows one to infer the Larmor phase and thus an average Larmor frequency over the sensing time~\cite{eto}. Smaller sensors are of course beneficial for improving spatial resolution, but absent increases in density and deleterious effects thereof, such sensors consequently have fewer spins and the precision of phase measurement is reduced. For large atom count BECs, the uncertainty of this Larmor phase measurement can be less than $\unit[10]{mrad}$~\cite{vengalattore,Hardman}, sensing magnetic field amplitude with several orders greater precision than classical sensors of comparable volume~\cite{mitchell}. For a given uncertainty in this phase, the uncertainty in the measured field $\delta B$ scales with the reciprocal of sensing time~$T$~\cite{Budker}. This favourable time-scaling makes long sensing times key to improving the sensitivity $\delta B$. 

The phase imputed from a single projective measurement of the final hyperfine populations is necessarily ambiguous due to the non-unique value of the Larmor phase corresponding to a final population difference. A given population difference is only unique within a domain $\theta \in [-\pi/2, \pi/2]$, which is termed the principal Ramsey fringe~\cite{Degen,Shiga}, and thus inferring a measured field $B$ from the measured populations requires that $B$ lies within $\pm \pi/(2\gamma T)$ of a known value, where $\gamma$ is the gyromagnetic ratio. Any violation of this assumption, referred to here as `fringe hopping', will lead to erroneous phase extraction from the spin projection, and resultant gross errors in the calculated field several orders of magnitude above that imposed by the phase uncertainty. A mis-calibration of the field, or unintended drift, can lead to systematic misattribution of the fringe. More insidiously, we explain below that in the case of an unshielded magnetometer, prevalent levels of broadband magnetic field noise induce fringe-hopping errors in Ramsey measurements with long interrogation time. 

Here we report unambiguous and precise magnetometry from a single, second-long interrogation of an ultracold atomic cloud. This cloud is produced in high-vacuum by truncating the evaporation process in our BEC apparatus, based on a hybrid magnetic and optical trapping configuration~\cite{Lin} and built across a $\unit[3\times 1]{m}$ optical table, the design of which is provided in detail in the previous work of our group~\cite{Wood2}. From a continuous measurement of spin projection we retrieve a conclusive measurement of dc magnetic field strength to nine significant figures, as well as a band-limited ac measurement of the field, from a sensing volume of $\unit[310 \times 10^{-15}]{m^3}$. A Faraday light-matter interface continuously detects atomic spin projection, enabling a Hilbert transform process to reconstruct the Larmor phase evolution. Hilbert-based instantaneous phase extraction has been applied to magnetometers based on NMR~\cite{Hong} and optical magnetometers based on shielded warm atomic vapours~\cite{Wilson}, but has thus far been applied only to large sensors, read out with correspondingly high signal-to-noise ratio (SNR). By contrast, our microscale sensor has a factor $10^5$ fewer spin quanta, and thus operates in a regime approaching the atom standard quantum limit (SQL).

\section{Measurement in noisy environments}

To consider the effect of magnetic noise on our measurement, we will begin from first principles by defining the relevant Hamiltonian. A full outline of our three-level spin-one Hamiltonian including relevant couplings is given in Appendix A, but we will lay out an illustrative two-level system here. Choosing the quantization axis to be along the magnetic field B(t)$\hat{z}$, the Zeeman Hamiltonian is $\hat{H}(t) = \gamma B(t) \frac{\hbar}{2} \hat{\sigma}_z$, where $\hat{\sigma}_z$ is the Pauli-Z operator. Considering the time-dependent Schrödinger equation, $-i \hbar \partial_t \psi = \hat{H}(t) \psi$, we may in general find the solution $\psi(t)$ through the Magnus expansion \cite{blanes}, $\psi(t) = \mathrm{exp}({\hat{\Omega}(t)}) \psi(0)$, where $\hat{\Omega}(t)$ is given by
\begin{multline}
    \hat{\Omega}(t) = \frac{1}{i\hbar} \int_0^{t} \hat{H}(t_1) dt_1 \\
+ \frac{1}{2(i\hbar)^2} \int_0^{t} \int_0^{t_1} \left[\hat{H}(t_1),\hat{H}(t_2)\right] dt_1 dt_2 + ...
\end{multline} 
Of note is the fact that all terms but the first are zero for any Hamiltonian which commutes with itself at all t. For a spin-half system subject to the previously established Hamiltonian, the spin thus evolves as 
\begin{equation}
    \psi(t) = \begin{bmatrix}
\exp\left(\frac{-i}{2} {\int_0^{t} \omega(\tau) d\tau}\right) & 0 \\
0 & \exp\left(\frac{i}{2} {\int_0^{t}\omega(\tau) d\tau}\right)
\end{bmatrix} \psi(0),
\label{eq:magnus}
\end{equation}
where we define the Larmor frequency $\omega(t) = \gamma B(t)$. This simplification applies for magnetic fields which change in magnitude, but not those which change direction, which can instead generate non-adiabatic Landau-Zener transitions~\cite{Bounds}. The result of this evolution is the creation of a phase difference between the spin states, the Larmor phase, given by 
\begin{equation}
    \phi(t) = \int_{0}^t \omega(\tau) d\tau = \int_{0}^t \gamma B(\tau) d\tau.
    \label{eq:phasedef}
\end{equation} 
In a Ramsey measurement, after some duration $T$ of free evolution, the final phase is inferred by a rotation and projection of the spin onto the quantization axis, yielding $\phi(T) = \mathrm{arcsin}(\langle \hat{\sigma}_z(T) \rangle)$. In the case of a linearly sensitive Ramsey measurement, in which the readout rotation is in quadrature with the initial rotation to maximize sensitivity to small field changes~\cite{taylor}, the Ramsey phase exhibits wrapping for $|\phi| > \frac{\pi}{2}$, and as such the measured $\langle \hat{\sigma}_z \rangle$ no longer specifies the correct $\phi$. Any such measurement where this phase cannot be constrained necessarily produces an ambiguous estimate~\cite{Waldherr}.

In this work, we consider a stationary field $B(t) = B_0 + \epsilon_{B}(t)$, where $\epsilon_{B}(t)$ is an additive white Gaussian noise process with variance $\sigma_B^2$ and power spectral density $S_\text{BB}$, and $B_0$ is the static field magnitude we seek to measure. For any finite measurement time, the average $\bar{B}$ of the noise is almost never zero. Considering a measurement over an interrogation time $\tau$ as a finite sample from this distribution, this average $\bar{B}$ will itself take on a Gaussian distribution centred on zero, with standard deviation 
\begin{equation}
    \sigma_{\bar{B}} = \sqrt{\frac{S_\text{BB}}{2\tau}}. 
    \label{eq:averagefieldspread}
\end{equation} 
Thus for a Ramsey measurement of duration $\tau$, the phase shift as a result of the magnetic noise $\epsilon_{B}(t)$ will take on a Gaussian distribution centered on zero with standard deviation
\begin{equation}
    \sigma_{\phi} = \gamma \sqrt{\frac{S_\text{BB}\tau}{2}}. 
    \label{eq:phasespread}
\end{equation}
Consequently, the noise spectral density of the field imposes a fundamental limit on the Ramsey interrogation time for unambiguous magnetometry, when measuring with an unshielded magnetometer.

From Eq.~\eqref{eq:phasespread}, we establish a critical time for $n$-sigma confidence of not exceeding $\phi = \pi/2$, 
\begin{equation}
    \tau_c(n) = \frac{\pi^2}{2 n^2 \gamma^2 S_\text{BB}}.
    \label{eq:criticaltime}
\end{equation}
In a typical laboratory noise environment, the noise amplitude spectral density $s_\mathrm{B} = \sqrt{S_\text{BB}}$ is approximately $100\;\mathrm{pT/\sqrt{Hz}}$~\cite{Woltgens}, and higher in close proximity to high-current power supplies. A $^{87}$Rb magnetometer in such an environment has 2-sigma critical time $\tau_c(2) = \unit[64]{ms}$, after which more than 5\% of Ramsey measurements will fringe hop, leading to gross error of order $1/\gamma \tau$ in inferring the static field $B_0$. This error rate quickly invalidates central value estimation even from an arbitrarily large number of measurements. Considering that the coherence time of ultracold atomic magnetometers can exceed several seconds~\cite{Deutsch}, this makes Ramsey measurement fundamentally unsuited to unambiguous long-period measurements on such mangetometers in the absence of shielding from external noise. Shielding ultracold atomic magnetometers has proven formidably challenging given optical access requirements and is incompatible with applications of these magnetometers to mapping fields on macroscopic samples with inherent magnetic noise sources. It is this objective of unshielded magnetometry that motivates the development of our continuous phase reconstruction protocol. 

\section{Apparatus}

To realize the protocol outlined in this work, we cool $^{87}$Rb atoms to $\unit[1]{\mu K}$, selectively catching atoms in the $|F,m_F\rangle \, = \, |1,-1\rangle$ state to produce an optically trapped cloud with a radius of $\unit[68]{\mu m}$ containing $1.6 \times 10^6$ atoms. Three orthogonal pairs of coils provide control over the local magnetic field. The $m_F$ states are energetically split by a $\unit[\sim 10^{-4}]{\mu T}$ axial bias field in an arbitrarily chosen axis, with transitions between these states controlled by resonant rf radiation. Stern-Gerlach measurement of the atomic spin is achieved by releasing the trap, applying a strong magnetic field gradient, and performing time-of-flight absorption imaging of the constituent $m_F$ populations. 

The atomic spin may alternatively be continuously measured by means of an off-resonant Faraday probe beam at the ``magic wavelength`` of $\lambda_\mathrm{mag} = \unit[790.03]{nm}$~\cite{Jasperse}, between the D1 and D2 lines of $^{87}$Rb, focused to a 150 $\mathrm{\mu m}$ waist to provide approximately constant illumination intensity across the atomic cloud. Due to the optical Faraday effect, this probe beam undergoes a polarization rotation proportional to the total spin projection onto the propagation direction of the probe, and this rotation is then measured by a balanced polarimeter as shown in Fig.~\ref{fig:apparatus}. The amplitude of this polarimeter signal depends on probe power, atom count, and beam alignment, and is typically normalised by the maximum recorded value. Typically, we use a bias field of order $\unit[100]{\mu T}$ perpendicular to the probe beam wave vector, maximizing the measurement of the transverse spin projection along the beam axis, which oscillates at the Larmor frequency. The quadratic Zeeman shift is suppressed by the ac Zeeman shift of an off-resonant microwave field at $\omega_\mathrm{mw}$, detuned to the red of the $|1,0 \rangle \leftrightarrow |2,0 \rangle$ clock transition, similar to the use of applied tensor light shifts to suppress the quadratic Zeeman shift in thermal atomic magnetometers~\cite{Jensen,Chalupczak2}. This reduces the quadratic shift to a residual value of approximately $\unit[0.1]{Hz}$, preventing the periodic decay and revival of the Larmor signal over our measurement lifetime~\cite{Jasperse}. The resultant polarimeter signal decays exponentially with lifetime \unit[530]{ms} and remains detectable well past one second. 

In order to perform dc and ac measurement of the local magnetic field, we produce a long-duration Faraday polarimeter recording of the free induction decay (FID) of Larmor precession. The initial spin eigenstate does not evolve, and no polarimeter signal is recorded. The photodetector and probe beam are switched on sequentially to characterize the electronic and optical noise spectra. The spin is then tipped into the transverse plane by a $\unit[13]{\mu s}$ duration resonant $\pi/2$ pulse, initiating precession at Larmor frequency $\omega(t) \approx 2\pi\times\unit[604]{kHz}$. Any variation of the magnitude of the bias field appears as modulation of the Larmor frequency, making the system linearly sensitive to small changes in field in the axis of our applied bias field. While it is possible to estimate the instantaneous Larmor frequency as the numerical derivative of the instantaneous phase of the polarimeter signal, such a time-varying frequency estimate can only be averaged incoherently. In this work, we show that by reconstructing and working with the phase directly, we preserve the coherence of our measurement and exploit the favourable time scaling to achieve superior sensitivity using the long coherence time of our quantum sensor. 

\section{Larmor Phase Reconstruction} 

We now consider the specific problem of estimating a stationary frequency $\omega$ in the presence of noise by linear regression to $\phi(t) = \omega t$~\cite{Fowler}. The reconstruction process begins with the polarimeter voltage signal 
\begin{equation}
    V(t) = A(t) \sin\left(\phi(t) + \phi_0\right) + \epsilon(t),
    \label{eq:voltagesignal}
\end{equation}
where $A(t)$ is the instantaneous amplitude, $\phi(t)$ is the instantaneous Larmor phase, $\epsilon(t)$ is an additive white Gaussian noise process with variance $\sigma^2$ band-limited at the Nyquist frequency corresponding to a sampling rate $f_s$ = 5 \unit[MSa/s]. The reference phase $\phi_0$ accounts for rf, atomic, and electro-optical delays between the local oscillator and digitisation of the polarimeter signal. The SNR at any given time is defined as the full-bandwidth SNR, given by $\mathrm{SNR} = A_{\mathrm{rms}}^2/\sigma^2 = A^2/2\sigma^2$, as is standard in relevant signal processing literature~\cite{Fowler}, not to be confused with the post-filter SNR defined in certain other literature on Faraday polarimetry~\cite{smith, Jasperse}. The polarimeter signal is then filtered with a zero-phase sixth-order Butterworth bandpass filter around the Larmor frequency, giving
\begin{equation}
    V_{pb}(t) =  A(t) \sin\left(\phi(t) + \phi_0\right) + \epsilon_{pb}(t),
    \label{eq:filteredsignal}
\end{equation}
where $A(t)$ and $\phi(t)$ are unchanged under the assumption that the signal power is entirely contained within the passband, which will be verified in experiment in Sec. VI. As this filtering is applied post-experiment, we can make use of a non-causal zero-phase filter which does not significantly alter the signal phase near the centre of the band. The purpose of this pre-filtering is to reduce the threshold SNR required for achieving the Cramer-Rao Lower Bound (CRLB) for phase-based frequency estimation by least-squares regression \cite{Fowler, Kim}, as will be described in Sec. VII.
This has a typical threshold $\mathrm{SNR}_{\mathrm{thr}}$ of 6 dB. 
Other estimators have been considered~\cite{Luise, Fitz, zhao}, but their computational complexity makes them ill-suited for applications with high sample counts. 

Considering a passband with equivalent noise bandwidth $\Delta f_{\mathrm{pb}}$, the threshold SNR is reduced by $\Delta \mathrm{SNR}_{\mathrm{thr}} = 10\log_{10}(2\Delta f_{\mathrm{pb}}/f_s)$ dB, and the maximum allowable equivalent noise bandwidth to maintain the CRLB is given by 
\begin{equation}
    \Delta f_{\mathrm{pb}} = \frac{f_s}{2} \cdot \frac{\mathrm{SNR}}{10^{3/5}}.
    \label{eq:passbandwidth}
\end{equation}
The passband of the filter transmits only a small fraction of the input power, which is overwhelmingly dominated by photon shot noise. The frequency of the Larmor precession can be easily identified in the polarimeter power spectrum by comparison to the pre-tip  spectrum outlined in Sec. II, allowing us to correctly center the passband. The band-filtered signal $V_\text{pb}$ is then used to produce an analytic signal representation of the polarimeter measurement,
\begin{equation}
    V_{a}(t) = V_{pb}(t) + i \mathcal{H}\left[{V}_\mathrm{pb}\right](t),
    \label{eq:analytic signal}
\end{equation}
where $\mathcal{H}\left[\cdot\right]$ represents the Hilbert transform. The details of the Hilbert transform and analytic signal representation are well-presented elsewhere~\cite{bracewell}; its utility in this application is the conversion of a real signal with conjugate positive and negative frequency components to a complex-valued signal with only positive frequency components and a well-defined phase. Importantly, so long as the Larmor signal power is entirely captured within the passband, no signal information is lost or distorted in the conversion to the filtered analytic signal representation $V_{a}(t)$. 

\begin{figure}[t]
\includegraphics[scale=0.305]{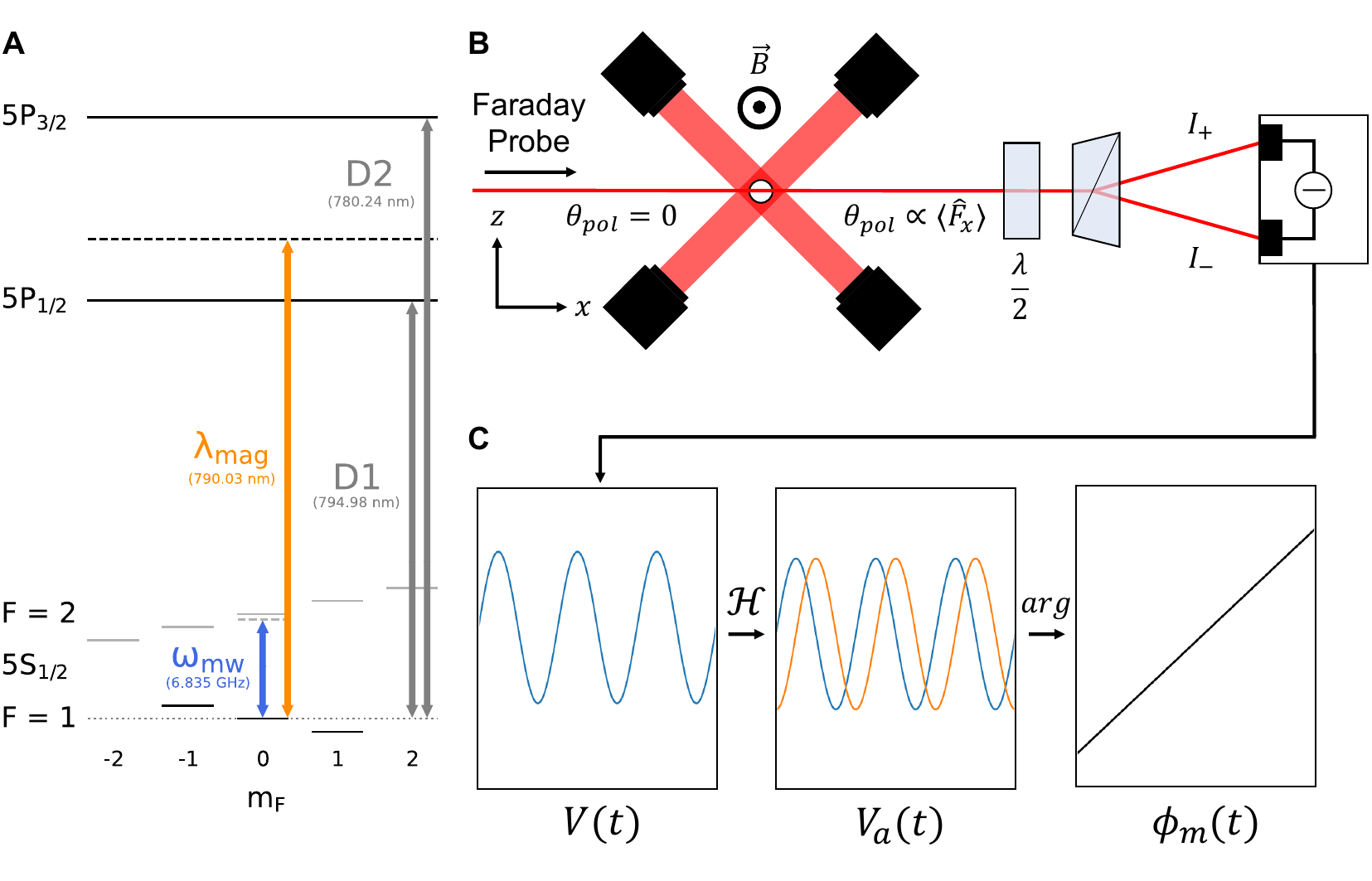}
\caption{Level structure, precession measurement and Larmor phase reconstruction: The spin dynamics of the ultracold atomic cloud in the $F = 1$ hyperfine ground state are continuously probed by a far off-resonant probe at the magic wavelength $\lambda_\mathrm{mag}$, with the quadratic shift nulled by off-resonant microwave dressing at frequency $\omega_\mathrm{mw}$ (a). The Faraday spin-light interface couples Larmor precession in a field $B_0 \hat{y}$ to the polarization rotation angle of a far-detuned probe beam focused onto the atoms (b). A balanced polarimeter (half-wave plate, Wollaston prism and differential photodetector) transduces the Larmor signal to a voltage $V(t)$ digitized at $\unit[5]{MSa/s}$ and 16 bits. The analytic signal representation $V_a(t)$ is formed using the Hilbert transform $\mathcal{H}$, the unwrapped argument of which is the phase $\phi_m(t)$, our estimate of the Larmor phase $\phi(t)$ (c).}
\label{fig:apparatus}
\end{figure}

The analytic representation is then recast in polar form as $V_{a}(t) = V_m(t) e^{i \phi_m(t)}$, where $V_m(t) = |V_{a}(t)|$ is the instantaneous amplitude envelope and $\phi_m(t) = \arg(V_{a}(t))$ is the (wrapped) instantaneous phase of the signal. As our Nyquist frequency well exceeds our Larmor frequency, this phase may be trivially unwrapped using NumPy's unwrap routine \cite{harris}. This unwrapped instantaneous phase is a measurement of the relative Larmor phase, and it is the evolution of this phase as governed by Eq.~\eqref{eq:phasedef} that allows us to perform dc and ac magnetometry with the system. The measurement and reconstruction workflow of our magnetometer is shown in Fig.\ref{fig:apparatus}. 

\section{AC magnetometry}

\begin{figure}[t]
\includegraphics[scale=0.46]{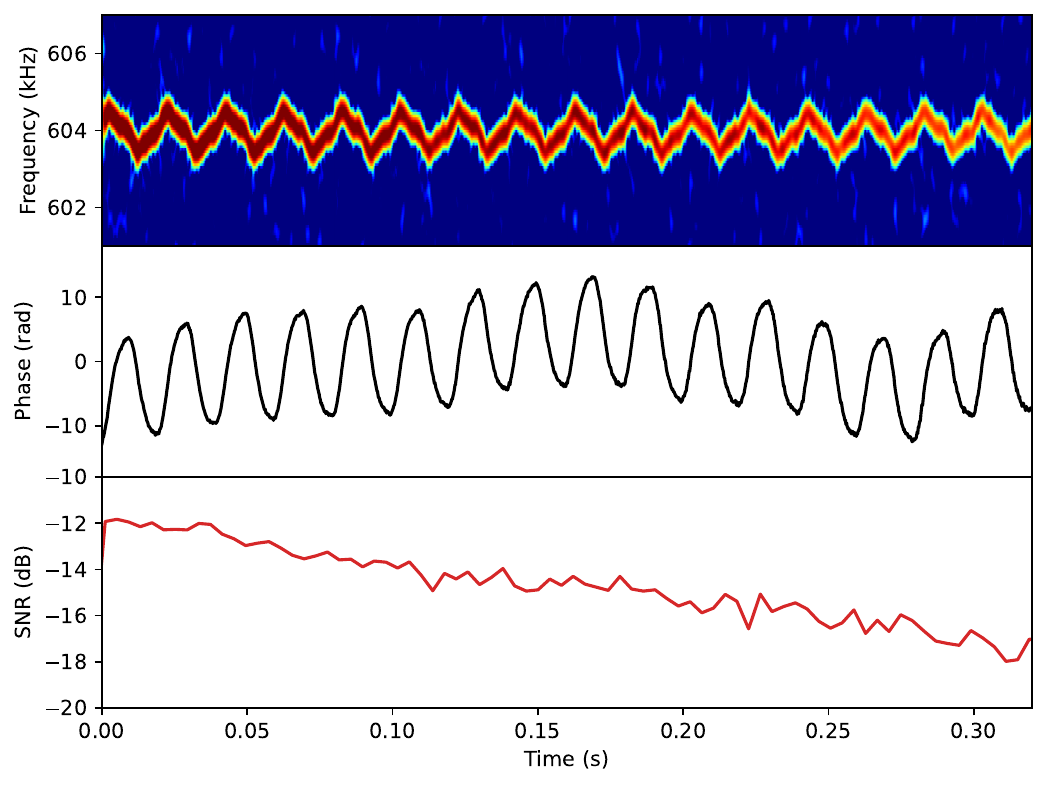}
\caption{{Continuous spin measurement: the polarimeter signal, a frequency-modulated sinusoid as seen in its spectrogram (top) is transformed into an analytic signal representation via the Hilbert transform. The phase function (middle) extracted from this is shown with the linear term removed to make plain the frequency modulation. The full-bandwidth SNR of the polarimeter signal (bottom) decays exponentially over the reconstruction time of \unit[320]{ms}, due to off-resonant scattering and spin dephasing.}}
\label{fig:noisetriple}
\end{figure}

While our field model thus far has considered only a constant magnetic field with additive white Gaussian noise, there are additional field contributions that must be considered in real environments. As an unshielded sensor, our apparatus is sensitive to magnetic fields from adjacent electronics. In our laboratory, low-frequency magnetic interference is dominated by line-synchronous oscillations at the power line fundamental of $\unit[50]{Hz}$, and its odd harmonics. In a typical environment, the fundamental component has amplitude of order $\unit[100]{nT}$ ~\cite{Woltgens}, with the \unit[150]{Hz} harmonic of order $\unit[30]{nT}$, and higher harmonics correspondingly weaker. A common technique to mitigate this interference is to synchronize the measurement to the line cycle and limit interrogation times to be much shorter than the line period $\tau \ll \unit[20]{ms}$. This greatly limits the sensitivity of the magnetometer in cases such as ours where the coherence time exceeds one second. When interrogating for $\tau \gg \unit[20]{ms}$, we resolve this interference as frequency modulation of the Larmor carrier frequency, allowing us to extract interference amplitudes from the phase modulation. The reconstruction of the Larmor phase over a duration of 320 ms, or 16 line cycles, in the unshielded laboratory environment is shown in Fig.~\ref{fig:noisetriple}.

From our \unit[320]{ms} reconstruction, we extract the RMS amplitudes of the \unit[50]{Hz}, \unit[150]{Hz}, and \unit[250]{Hz} field harmonic components of \unit[41.92(3)]{nT}, \unit[10.88(9)]{nT}, and \unit[2.0(1)]{nT} respectively by least-squares regression of a harmonic interference model to the phase function. The ac sensitivity is quantified and discussed in Sec. VII. The phase reconstruction duration is limited by the declining polarimeter SNR as scattering of probe light removes atoms from the cloud. In theory, the reconstruction SNR can be improved by a reduction in sensor bandwidth through tightening the passband, initially chosen to be $\unit[5]{kHz}$; however, there exists a minimum bandwidth below which we impinge on the signal band. The amplitude of the harmonic components determines the frequency deviation of the Larmor signal, and sets this minimum bandwidth, estimated at $\unit[2.5]{kHz}$ peak-to-peak on previous work with this apparatus~\cite{anderson} using Carson's rule, which provides a good approximation of the spectral support of wide band frequency-modulated signals~\cite{Carson}. Strong field fluctuations lead to large variations in Larmor frequency, requiring a wider pass bandwidth which admits more photon shot noise, and thus bringing forward the time by which the polarimeter SNR is insufficient for accurate phase reconstruction. 

The corresponding minimum SNR for least-squares phase estimation considering our initially chosen pass bandwidth of $\unit[5]{kHz}$ is $\unit[-21]{dB}$ as provided by \eqref{eq:passbandwidth}. This limits the reconstruction duration in the case of the measurement shown in Fig.~\ref{fig:noisetriple} to approximately \unit[500]{ms} before we can no longer achieve the CRLB through least-squares estimation. Additionally, once the SNR has fallen significantly below this threshold, phase reconstruction may fail entirely due to phase-unwrapping errors, leading to discontinuities in the reconstruction. For this reason, it is imperative to reduce frequency modulation from magnetic interference to allow for the narrowest possible passband, and hence the longest possible measurement time. 

\section{Feed-forward noise cancellation}

As magnetic interference from electrical equipment is a common problem in metrology experiments, there already exists a body of literature discussing methods of suppressing it. These range from simple feed-forward control cancelling periodic noise using prior field recordings~\cite{eto,merkel} to complex feedback mechanisms combining measurements from several secondary sensors in the vicinity of the primary sensor~\cite{Platzek, xu2}. While feedback control is effective against periodic \emph{and} non-periodic noise, the aforementioned secondary sensor methods necessarily measure the field at some distance from the atoms, and struggle to neutralise the spatially-varying interference arising from multiple sources. On the other hand, using the atomic sensor as the input for feedback control~\cite{Zhang} necessarily sacrifices limited quantum resources. While capable of providing strong suppression of ac magnetic interference in stable environments, feed-forward noise cancellation in ultracold atoms to date has relied on performing a long series of calibration measurements~\cite{eto, smith3}, making the system vulnerable to short-term drifts in interference amplitude and phase. 

\begin{figure}[t]
\includegraphics[scale=0.46]{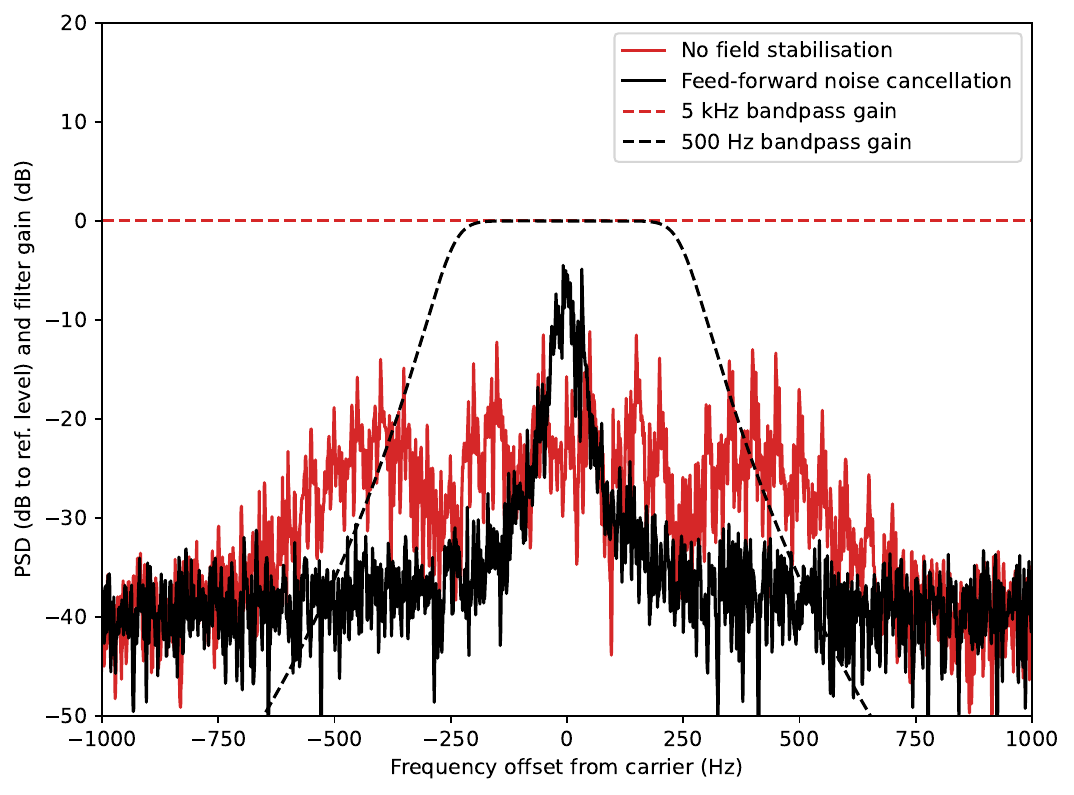} 
\caption{{Feed-forward noise cancellation reduces the bandwidth occupied by the Larmor signal: due to harmonic interference, the power spectral density (PSD) of the Larmor signal in a non-stabilised environment (red) has significant signal power at deviations up to $\unit[750]{Hz}$ from the carrier frequency. With feed-forward noise cancellation (black) this deviation is reduced by almost an order of magnitude, permitting reduction of the filter bandwidth from $\unit[5]{kHz}$ (dashed red) to $\unit[500]{Hz}$ (dashed black). The PSDs share the same reference level. Note that the 5 kHz bandpass filter gain rolls off by only 0.3 dB across the displayed range.}}
\label{fig:powerspectrum}
\end{figure}

Acknowledging that the dominant contribution to the modulation is the quasi-stationary line-synchronous interference, we may use our measurement of the local magnetic environment to perform feed-forward noise cancellation by producing a complementary ac field. This complementary field is synchronised to the electrical line phase at the beginning of each measurement by an external trigger, and is produced by small, single-turn shim coils with a diameter of $\unit[6]{cm}$, fastened to the interior surface of a cylindrical polyvinyl chloride mount and placed co-axially with the bias coils. The coil currents are controlled by an analog output card which feeds a voltage-controlled current supply, with an overall system time constant of $\unit[40]{\mu s}$, allowing arbitrary ac magnetic fields to be produced with a maximum amplitude of $\unit[6.61(3)]{\mu T}$ and bandwidth of $\unit[10]{kHz}$. The amplitudes and phases of the modulation terms extracted from the reconstructed phase evolution in our unshielded atomic cloud (Fig.~\ref{fig:noisetriple} (middle)) calibrate the noise cancellation. The effects of this noise suppression are readily apparent in the power spectrum of the polarimeter signal: where the Larmor resonance is initially spread over a wide range of frequencies, suppression of the line-synchronous interference results in a narrow resonance with no defined sidebands. This allows us to tighten the bandpass filter from $\unit[5]{kHz}$ to $\unit[500]{Hz}$ while keeping all of the resonance within the flat range of the passband. The polarimeter signal power spectra and the relevant bandpass filter gains are shown in Fig.~\ref{fig:powerspectrum}.

\begin{figure}
\includegraphics[scale=0.46]{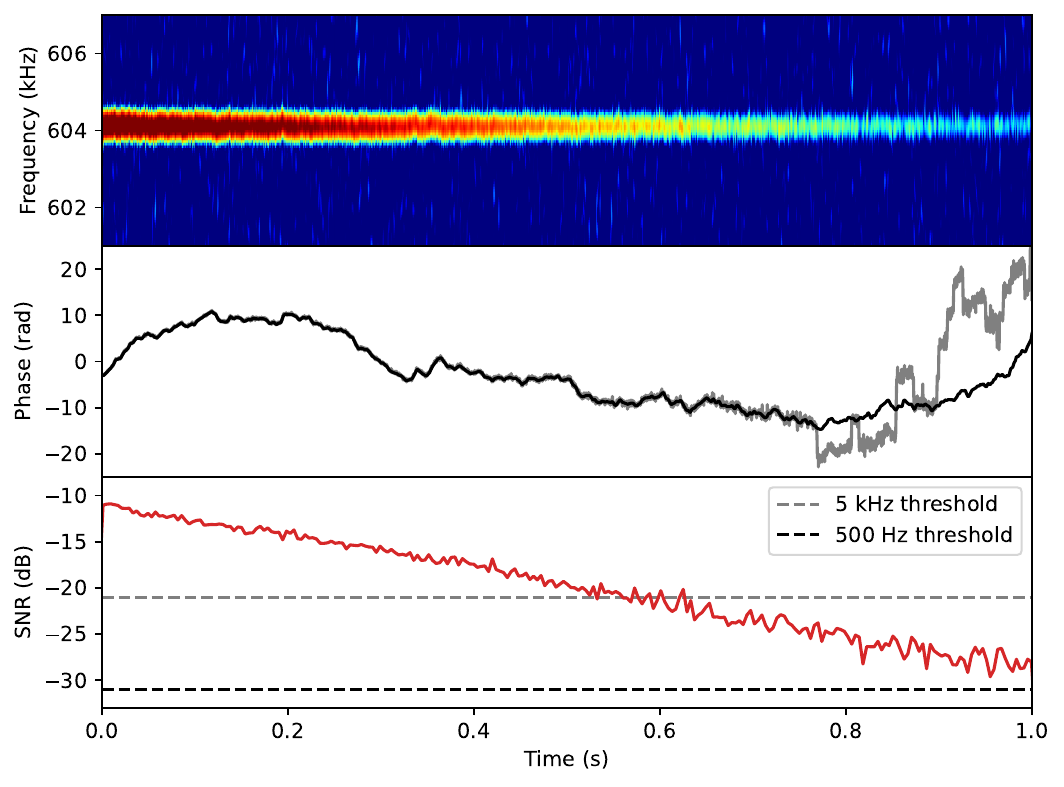}
\caption{{Continuous spin measurement with feed-forward noise cancellation: the  spectrogram (top) now exhibits a stable carrier frequency without visible frequency modulation. The phase function residuals (middle) demonstrate random walk behaviour with discontinuities appearing at later times in the reconstruction due to phase-wrapping errors when using the wider bandwidth of \unit[5]{kHz} (grey), but successful reconstruction at the tighter \unit[500]{Hz} bandwidth (black). Frequency measurement by phase estimation achieves the Cramer-Rao bound so long as the SNR remains above the threshold for the respective bandwidths (bottom).}}
\label{fig:antinoisetriple}
\end{figure}

Fig.~\ref{fig:antinoisetriple} shows the polarimeter signal, reconstructed phase, and SNR, now with feed-forward noise cancellation enabled, over almost one second of interrogation time. In comparison with Fig.~\ref{fig:noisetriple}, the line harmonics are now imperceptible in the spectrogram, and indeed there is no visible harmonic fluctuation in the residual phase. 
This reduction of the pass bandwidth from $\unit[5]{kHz}$ to $\unit[500]{Hz}$ permits the reconstruction time to be extended out to the full $\unit[1000]{ms}$ while maintaining sufficient SNR to achieve the CRLB. 
As a result, the phase retrieved from the narrower filter (black trace in Fig.~\ref{fig:antinoisetriple} (middle)) shows no sign of phase unwrapping errors across the full reconstruction time, whereas the phase retrieved from the original filter (gray trace) manifests a series of such errors once the SNR falls significantly below threshold beyond approximately $\unit[750]{ms}$.

The efficacy of feed-forward noise cancellation depends almost entirely on the stability of the local magnetic environment. Changes in the power draw of high-current devices (such as motors, amplifiers, and HVAC equipment) cause shifts in the phase and amplitude of the line harmonics, leading to reduced noise cancellation until a new ac magnetometry calibration measurement is performed. The measurement and reconstruction process, from the beginning of trap loading to completion of post-processing, takes less than one minute, allowing rapid re-calibration of noise cancellation in the event of a change in the magnetic environment. In practice, a calibration measurement is performed daily to account for changes in the position or current draw of local equipment. 

\begin{figure}[t]
\includegraphics[scale=0.46]{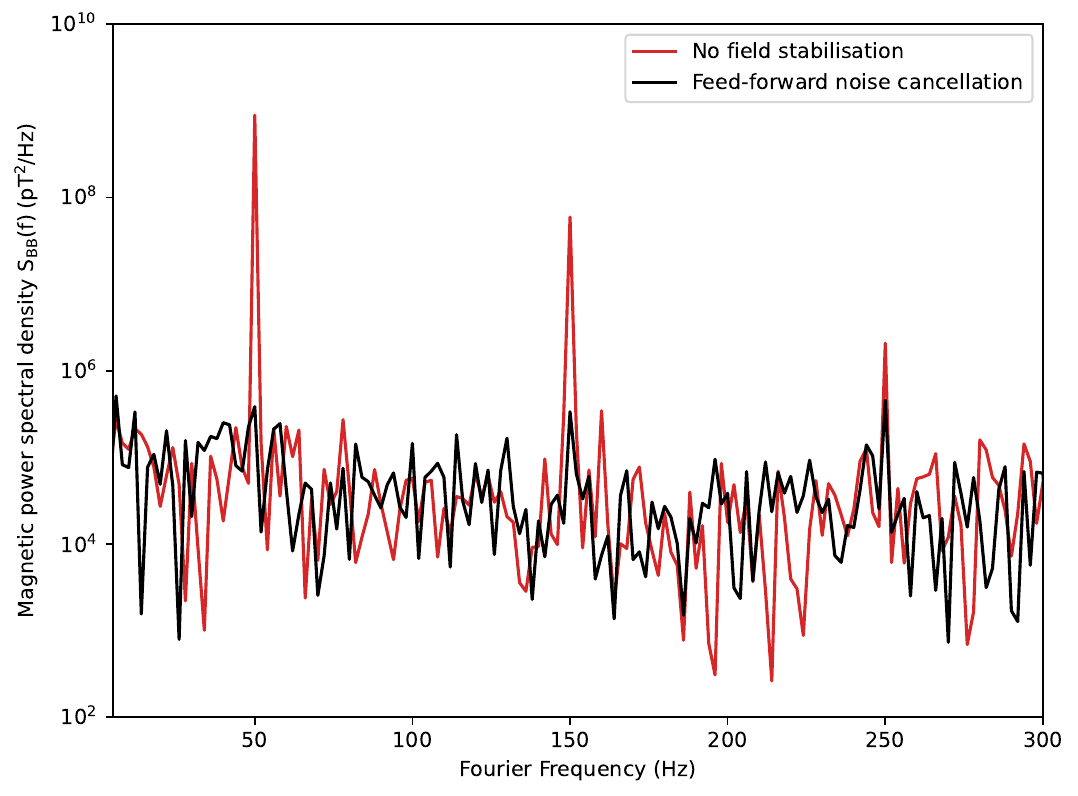} 
\caption{{Feed-forward noise cancellation eliminates line harmonics from the magnetic noise spectrum: the magnetic power spectral density is reconstructed from single-shot Faraday polarimeter signals with (black) and without (red) feed-forward noise cancellation. The Fourier-limited frequency resolution of this spectrum is $\unit[2]{Hz}$.}}
\label{fig:fieldspectrum}
\end{figure}

We quantify the field stability by analysing the magnetic power spectral density $S_{\text{BB}}(f)$ in the sub-300 Hz band, estimated by periodogram of the derivative of the Larmor phase as per Eq.~\eqref{eq:phasedef}. Additionally, we define the equivalent RMS magnetic noise amplitude
\begin{equation}
\delta B_\text{noise} = \sqrt{\int_0^{f_\text{max}} S_\text{BB} \, df},
\label{eq:fielddeviationintegral}
\end{equation}
which for the laboratory field is measured to be~\unit[44.4]{nT} for a $f_\mathrm{max}$ of \unit[300]{Hz}. Feed-forward noise cancellation reduces this by a full order of magnitude to $\delta B_\text{noise}=\unit[4.4]{nT}$ representing a \unit[20.1]{dB} reduction in  interference in the sub-300 Hz band; this accounts for the significant reduction in the bandwidth occupied by the Larmor signal shown in Fig.~\ref{fig:powerspectrum}. Fig.~\ref{fig:fieldspectrum} shows the measured magnetic noise power spectral density $S_\text{BB}$, confirming that the laboratory environment is dominated by the 50, 150, and \unit[250]{Hz} line interference harmonics. 

Making use of feed-forward noise cancellation, the line-synchronous interference is suppressed to become only marginally resolvable against the white noise background of $\unit[250]{pT/\sqrt{Hz}}$. Owing to drifts in the grid frequency of up to $\unit[100]{mHz}$~\cite{aemc}, feed-forward noise cancellation is most effective for the fundamental frequency of $\unit[50]{Hz}$, and less effective for higher harmonics, for which this frequency error is multiplied. It is important to clarify that this figure of $s_\mathrm{B} = \unit[250]{pT/\sqrt{Hz}}$ is not a sensitivity, and merely characterises the stability of the background field. This magnetic noise spectral density is marginally higher than typical values reported in a laboratory environment~\cite{Woltgens}, and is likely a result of our use of high-current equipment in relatively close proximity to the apparatus. The $2 \sigma$ critical time predicted by Eq.~\eqref{eq:criticaltime} for this magnetic noise spectral density is only \unit[10]{ms}, limiting Ramsey measurement duration to a small fraction of the Zeeman coherence time of our ultracold atomic sensing platform.

\section{Phase Retrieval dc magnetometry}

\begin{figure}[t]
\includegraphics[scale=0.46]{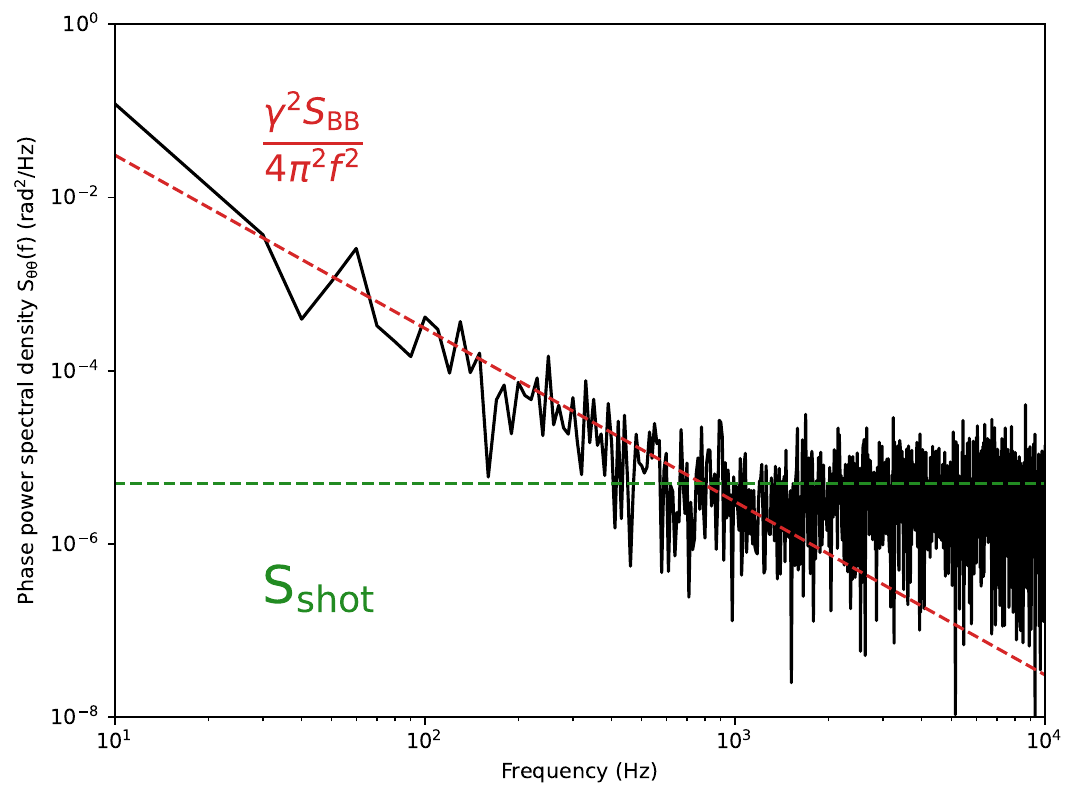}
\caption{{Phase power spectral density: the power spectrum of the Larmor phase residuals under feed-forward noise cancellation. Above the corner frequency of $\unit[800]{Hz}$, the noise spectrum is white as a result of photon shot noise. Below this frequency, the noise floor is dominated by red phase noise due to approximately white environmental magnetic field noise. The Fourier-limited frequency resolution of this spectrum is \unit[10]{Hz}.}}
\label{fig:phasespectrum}
\end{figure}

We realise dc magnetometry by performing a Faraday polarimetry measurement of the FID under feed-forward noise cancellation. The Larmor phase is then reconstructed using the same method as described in Sec. III. With the nearly-complete removal of the line-synchronous modulation, it is expected that the Larmor phase will increase linearly, and as such we now perform a least-squares regression of the reconstructed phase function $\phi_m(t)$ to the linear model $\phi_m(t) = \gamma B_\text{est} t + \phi_\text{est}$.
In the case of a truly static magnetic field, the phase residual is simply the white Gaussian photon and atom shot noise $\epsilon_{pb}(t)$ of the polarimeter voltage signal $V_{pb}(t)$ imputed as phase noise. However, in the case of an unshielded apparatus with appreciable fluctuations in the magnetic field, the phase noise has a colored spectrum, as shown in Fig.~\ref{fig:phasespectrum}.  This phase power spectral density can be decomposed into two terms, a colored noise term reflecting underlying noise in the measured magnetic field, and a white noise term as a result of imputing photon and atom shot noise as phase fluctuations, with the crossover frequency found at $\unit[800]{Hz}$. The measured photon shot noise background across the $\sim \unit[5]{MSa}$ of a single recording was found to be indistinguishable from additive white Gaussian noise by the Kolmogorov-Smirnov test ($D = 4.3 \times 10^{-4},\; p = 0.35$). 

As shown in Fig.~\ref{fig:fieldspectrum}, the magnetic power spectral density is approximately white under feed-forward noise cancellation, corresponding to the red noise in the phase power spectral density. Considering this, we define the model
\begin{equation}
S_{\phi \phi}(f) = \frac{\gamma^2 S_\text{BB}}{4 \pi^2 f^2} + S_\mathrm{shot},\\
\label{eq:phasenoisespectrum}
\end{equation}
where $S_\text{BB}$ is the magnetic power spectral density as defined previously, and $S_\mathrm{shot} = 2/(f_s \mathrm{SNR})$ for $f < \omega/(2\pi)$~\cite{Hong}. Thus, the phase noise power asymptotically increases as the Fourier limit frequency tends to zero, and would be unbounded in the limit $\tau \rightarrow \infty$. As interrogation time increases, commensurately more low-frequency noise contributes to the phase noise, becoming the dominant contribution for $\tau \gg \unit[1]{second}$. The sensitivity of the field estimate produced by this reconstruction is fundamentally limited by the regression error, resulting in a sensitivity
\begin{equation}
\delta B_\text{dc} = \frac{2 \,\delta\phi}{\gamma \tau^{\frac{3}{2}}}\sqrt{\frac{3}{f_s}},  \\
\label{eq:sensitivity}
\end{equation}
where $\delta\phi$ is the RMS amplitude of the phase residuals; the derivation of this expression is shown in Appendix B. Considering the noise spectra shown in Fig.~\ref{fig:phasespectrum} with a measurement duration of \unit[1]{second} and associated Fourier limit frequency of \unit[1]{Hz}, the majority of the total phase noise power is a result of detector noise, and as such we consider the detector noise limited regime identified in Appendix B. In this limit, in terms of $\mathrm{SNR}$, the dc sensitivity is given by
\begin{equation}
\delta B_\text{dc} = \frac{1}{\gamma \tau^{\frac{3}{2}}} \sqrt{\frac{12}{f_s \, \mathrm{SNR}}}. \\
\label{eq:sensitivitysnr}
\end{equation}
This metric illuminates the superior time-scaling achieved by a continuous phase-coherent measurement, motivating longer phase reconstruction duration. With the performance of our magnetometer characterised, we now perform precise dc magnetometry by continuous phase reconstruction under feed-forward noise cancellation. In an experimental run, a series of shots were performed measuring the $86.03(4) \; \mathrm{\mu T}$ static field of the laboratory environment. The high variability of this measured field is due to quasi-DC changes as a result of a rail line in proximity to our laboratory, as has been independently verified by a commercial flux-gate magnetometer. Taking a single shot in this series, we obtain Figures 3, 4, and 5, with a phase reconstruction spanning a sensing duration of $\tau_s = 1000 \mathrm{ms}$, in a sensing volume of $\unit[310 \times 10^{-15}]{m^3}$. From this reconstruction, we measure an average field of $\unit[86.0121261(4)] {\mu T}$, with an estimation-limited sensitivity of $\delta B_\text{dc} = \unit[380]{fT}$, and an inferred detector-limited sensitivity of $\delta B_\text{dc} = \unit[359(7)]{fT}$ calculated from Eq.~\eqref{eq:sensitivitysnr} using the weighted average SNR of \unit[-20.2(1)]{dB}, as described in Appendix B. This achieves superior sensitivity in such a volume when compared to other quantum platforms such as bulk nitrogen-vacancy centres in diamond, and approaches the sensitivity per unit volume reported in work on other (ambiguous) measurement protocols in ultracold atoms and BEC~\cite{mitchell}. This also marks a significant improvement over comparable frequency-based measurements using a Faraday measurement of ultracold atoms, which achieved a sensitivity of $\unit[141]{pT}$ in a sensing duration of $\unit[5]{ms}$ on a condensate of $\unit[3 \times 10^5]{atoms}$, for which the measurement duration was limited by ac modulation of the measured field rather than coherence lifetime~\cite{Jasperse}. With regards to ac sensing, the ac sensitivity can also be characterised by imputing an equivalent magnetic noise for a given amplitude of phase noise~\cite{Wilson}. Considering this, we find the expression 
\begin{equation}
\delta B_{ac}= \frac{2 \pi f}{\gamma\sqrt{f_s \,\tau \, \mathrm{SNR}}}. \\
\label{eq:acsensitivity}
\end{equation}
Using the initial SNR of $\unit[-11.1]{dB}$ achieved for measurement time $\tau < 10 \, \mathrm{ms}$ this corresponds to an initial bandwidth-normalized ac sensitivity of $\delta B_{ac} \sqrt{T} = \unit[230 \times f\;]{fT/\sqrt{Hz}}$. Importantly, unlike dc sensitivity, the bandwidth-normalised ac sensitivity of this sensor does not scale with increased measurement time. Naturally, to successfully measure an ac signal at frequency $f$, the pass bandwidth must exceed $2f$: as the probability of phase reconstruction errors increases with pass bandwidth, this establishes a maximum sensing bandwidth. Taking the maximum allowable passband width to be the same as that defined in \eqref{eq:analytic signal}, we find a maximum passband of $\unit[48.8]{kHz}$, and consequently a baseband ac sensing range of up to $\unit[24.4]{kHz}$. The ac bandwidth of the sensor can be improved by increasing the SNR of the Faraday signal, however this would require a brighter probe beam, reducing the scattering lifetime of the atoms and negatively impacting the dc sensitivity. 

The competitive dc and ac sensitivity of this sensor up to frequencies of several kHz, combined with its small active sensing volume, suggests that it will be able to meet the demands of microscale sensing applications in both biological and materials sciences, particularly if used as a scanning probe of magnetic structure~\cite{Yang}. The high-vacuum environment in which the ultracold atomic vapour is held does present a barrier to measurement of vacuum-incompatible samples, such as live cells, and as such the utility of our sensor is at present limited to measuring magnetically-active samples which can tolerate at least a low-vacuum environment. The minimally-destructive nature of the Faraday measurement also gives rise to the prospect of using it to supplement other measurements, such as a projective interferometry measurement, to provide a more complete picture of the spin dynamics over the sensing duration.  

\section{Conclusion}
We have developed a continuous phase reconstruction protocol on a microscale ultracold atomic sensor allowing precise, unambiguous dc measurement and ac measurement of magnetic field in a noisy environment. Measuring unshielded in Earth's field, the magnetometer resolves nine significant figures of the dc field magnitude in a single shot, undeterred by field drifts of up to several microtesla between shots. Additionally, we have demonstrated suppression of local magnetic interference by over 20 dB by feed-forward control. This sensor fulfills the requirements of rapid, calibration-free, unambiguous and unshielded magnetometry needed to realise magnetic micro-imaging of electrophysiological function, surface condensed-matter physics, and chemical structure on the cellular scale.

\section{Acknowledgements}
This work was supported by an Australian Government Research Training Program (RTP) scholarship, and funded by the Australian Research Council under Linkage Project number LP200100082. 

\appendix
\section{Spin-1 dynamics with microwave coupling}

As outlined in Sec. II, our alkali spinor is a three-level spin-1 system with free evolution governed by the diagonal Hamiltonian $\hat{H}(t)$. Owing to the precision of the measurement performed here, we consider this system in full generality using the Breit-Rabi equation for Zeeman energy eigenvalues in $J=1/2$ states~\cite{breit}, given by
\begin{multline}
E_{F,m}(B) = \frac{-E_{\mathrm{hfs}}}{2(2I+1)} + g_I \mu_B m B \\ \pm \frac{E_{\mathrm{hfs}}}{2}\sqrt{1+\frac{4 m x}{2 I +1} + x^2},
\label{eq:breitrabi}
\end{multline}
where $x = ({(g_J - g_I)\mu_B B})/E_{\mathrm{hfs}}$ is the dimensionless magnetic field magnitude, and $F=I\pm J$ defines the signs, so that the minus is adopted for our $F=1$ manifold, for $I=3/2$ in $^{87}$Rb. In the above equations, $\mu_B$ is the Bohr magneton, $E_{\mathrm{hfs}}$ is the ground-state hyperfine splitting, $g_J$ and $g_I$ are the fine structure, and nuclear Landé factors respectively, and $m$ is the magnetic quantum number. In the absence of any other couplings, this results in an $\mathrm{F} = 1$ manifold Hamiltonian given by 
\begin{equation}
    \hat{H} = \begin{bmatrix}
E_{1,+1}(B) & 0 & 0\\ 
0 & E_{1,0}(B) & 0 \\
0 & 0 & E_{1,-1}(B) \\
\end{bmatrix}, 
\label{eq:hamiltonian}
\end{equation}
where henceforth we suppress explicit time dependence. As the Hamiltonian is diagonal and thus commutes with itself at all times, we may define a Larmor frequency from the Magnus expansion as done in Sec. II, giving
\begin{equation}
\omega(t) = \frac{E_{1,+1}(B) - E_{1,-1}(B)}{2\hbar},
\label{eq:larmordef}
\end{equation}
and additionally define a quadratic Zeeman shift
\begin{equation}
q(t) = \frac{E_{1,+1}(B) + E_{1,-1}(B) - 2E_{1,0}(B)}{2\hbar}.
\label{eq:quadraticdef}
\end{equation}
Typically, the Breit-Rabi equation is only expanded to second order when considering weak-field dynamics, leading to
\begin{equation}
\omega = \frac{(5 g_i - g_j) \mu_B}{4\hbar} B + \mathcal{O}(B^3) \approx \gamma_0 B,
\label{eq:nonlinearlarmor}
\end{equation}
and
\begin{equation}
q = \frac{(g_i - g_j)^2 \pi \mu_B^2}{8 E_{\mathrm{hfs}} \hbar^2} B^2 + \mathcal{O}(B^4) \approx q_0 B^2,
\label{eq:nonlinearquadratic}
\end{equation}
where $\gamma = 2 \pi \times \unit[-7.02369]{GHz/T}$ and $q_0 = 2 \pi \times \unit[7.189]{GHz/T^2}$ are the zero-field gyromagnetic ratio and quadratic Zeeman shift respectively using known values~\cite{steck}. Outside the low-field limit, or where high precision is required, the cubic term in Eq.~\eqref{eq:nonlinearlarmor} can also be included, resulting in a $c_0 = \unit[44.24]{GHz/T^3}$ shift to the Larmor frequency. If this term is included, we may no longer decompose the Larmor frequency in terms of a constant gyromagnetic ratio $\gamma_0$. However, as the Larmor frequency remains a monotonic function of field, we may instead define a `running' gyromagnetic ratio $\gamma(B)$, defined such that
\begin{equation}
\omega(t) = \gamma(B) \times B.
\label{eq:linearlarmor}
\end{equation}
During `free' evolution of the spin, it interacts with two off-resonant radiation fields: the Faraday probe and the microwave driving. The scalar, vector, and tensor light-shifts as a result of the Faraday probe can be made arbitrarily small through use of a `magic-zero wavelength' and precise control of polarisation \cite{Jasperse}, leaving the off-resonant microwave coupling as the only additional term to consider in the Hamiltonian. A microwave source detuned from the $|1,0 \rangle$ and $|2,0 \rangle$ states cancels the quadratic Zeeman shift by inducing an ac Zeeman shift given by 
\begin{equation}
q_{mw, 0} = -\frac{\Omega^2_{\mathrm{mw}}}{4\Delta_\mathrm{mw}},
\label{eq:microwaveshift}
\end{equation}
where $\Omega_\mathrm{mw} \approx \unit[2\pi \times 6]{kHz}$ and $\Delta_\mathrm{mw} \approx \unit[2\pi \times 150]{kHz}$ are the microwave Rabi frequency, and detuning from the clock transition, respectively. In the limit $\omega \gg \Delta_{mw}$, this shift is only substantial for the $|1,0 \rangle \leftrightarrow |2,0 \rangle$ transition, however when we consider fields of order $\unit[100]{\mu T}$, the Zeeman shift is of the same order as the detuning, resulting in appreciable ac Zeeman shifts for the $m = \pm 1$ states. These shifts are given by 
\begin{equation}
q_{mw, \pm 1} = -\frac{\hbar \Omega^2_\mathrm{mw}}{4\hbar(\Delta_\mathrm{mw} - (E_{2,\pm 1} - E_{2,0}) + (E_{1, \pm 1} - E_{1,0}))},
\label{eq:pm1microwaveshift}
\end{equation}
with shifts on the $\pm 1$ states differing in both magnitude and direction. This leads to additional non-linear terms in $E_{F,m}(B)$, which must be included at the field magnitude and desired measurement precision in this work. As such, we define new eigenenergies including the microwave shifts, by
\begin{equation}
E'_{F,m}(B) = E_{F,m}(B) + q_{\mathrm{mw},m},
\label{eq:modifiedlevels}
\end{equation}
and recompute the Larmor frequency
\begin{equation}
\omega(t) = \frac{E'_{1,+1}(B) - E'_{1,-1}(B)}{2\hbar},
\label{eq:modifiedlarmor}
\end{equation}
and similarly the quadratic Zeeman shift
\begin{equation}
q(t) = \frac{E'_{1,+1}(B) + E'_{1,-1}(B) - 2E'_{1,0}(B)}{2\hbar}.
\label{eq:modifiedquadraticshift}
\end{equation}
Away from the resonance $\Delta_\text{mw} = \omega$, the Larmor frequency is a monotonic function of B, allowing us to take the decomposition 
\begin{equation}
\omega(t) = \gamma_{\Omega,\Delta}(B) \times B,
\label{eq:nonlinearlarmorevolution}
\end{equation}
where $\gamma_{\Omega,\Delta}(B)$ is the new running gyromagnetic ratio, which explicitly depends on the parameters of the microwave coupling as well as the field. The microwave frequency is locked to a precision reference, and contributes negligible error, and the Rabi frequency at the atoms is determined experimentally by nulling the total quadratic shift at a given B through a process of iterative optimisation. 

Thus, with known $\Omega_\text{mw}$, $\Delta_\text{mw}$, and relevant atomic parameters, frequency estimation can provide accurate field estimation at precisions which require inclusion of the non-linear elements of the Zeeman splitting. The running gyromagnetic ratio depends only weakly on B, and thus the decomposition used in Eq.~\eqref{eq:nonlinearlarmorevolution} illustrates the near-linear relation between field and Larmor frequency. 

\section{Sensitivity}

In order to quantify the dc sensitivity of a regression error-limited field estimate from phase reconstruction, we define the sensitivity as 

\begin{equation}
\delta B_\text{dc}  = \frac{\sigma_{\omega}}{\gamma}, \\
\label{eq:sensitivityfundamentaldef}
\end{equation}
where $\sigma_\omega$ is the standard error in the angular frequency estimate retrieved from least-squares regression of a time series [$t_i$, $\phi_i$] to a linear model over a sensing time $\tau$. The standard error in such an estimate is a well-established result in regression theory, given by 
\begin{equation}
\sigma_{\omega} = \sqrt{\frac{\sigma_r^2}{\sum_{i=0}^{N}(t_i - \bar{t})^2}}, \\
\label{eq:omegaspreadstats}
\end{equation}
with the standard deviation about the regression $\sigma_r$ equal to
\begin{equation}
\sigma_{r} = \sqrt{\frac{\sum_{i=0}^{N}(\phi_i - \hat{\phi}_i)^2}{N-2}}, \\
\label{eq:regressionspreadstats}
\end{equation}
where $\hat{\phi}_i$ is the regression estimate for $\phi_i$. The factor $\mathrm{N-2}$ is present due to the loss of two statistical degrees of freedom. In the limit of large samples,
\begin{equation}
\lim_{N\to\infty} \sigma_r = \delta\phi. \\
\label{eq:highnlimit}
\end{equation}
Additionally in this limit, we may take the continuum limit of the sum over $t_i$, giving
\begin{equation}
\sum_{i=0}^{N}(t_i - \bar{t})^2 = f_s \int_0^\tau (t-\frac{\tau}{2})^2 dt = f_s \frac{\tau^3}{12}. \\
\label{eq:continuumlimit}
\end{equation}
We may now rewrite Eq.~\eqref{eq:omegaspreadstats} as 
\begin{equation}
\sigma_{\omega} = \frac{2 \; \delta\phi}{\tau^{\frac{3}{2}}} \sqrt{\frac{3}{f_s}}, \\
\label{eq:omegaspreadsimplified}
\end{equation}
and equivalently rewrite Eq.~\eqref{eq:sensitivityfundamentaldef} as 
\begin{equation}
\delta B_\text{dc} = \frac{2 \; \delta\phi}{\gamma \tau^{\frac{3}{2}}}\sqrt{\frac{3}{f_s}}.  \\
\label{eq:phasedeltasensitivity}
\end{equation}
The residual phase variance $\delta\phi^2$ is the sum of variance arising from shot noise in the detector and variance as a result of noise in the actual field, 
\begin{equation}
\delta\phi^2 = \delta\phi_\text{shot}^2 + \delta\phi_\text{field}^2.  \\
\label{eq:phasenoisebreakdown}
\end{equation}
As previously discussed, shot noise is imputed as additive white Gaussian noise in our phase reconstruction, giving rise to an equivalent full-bandwidth phase variance of $\delta\phi_\text{shot}^2 = 1/\text{SNR}$~\cite{Hong}, while the field phase variance is given by
\begin{equation}
\delta\phi_\text{field}^2 = \int_{\frac{1}{\tau}}^\infty \frac{\gamma^2 S_\text{BB}}{4 \pi^2 f^2} df = \frac{\gamma^2 S_\text{BB} \tau}{4 \pi^2}.  \\
\label{eq:fieldphasenoise}
\end{equation}
Thus, 
\begin{equation}
\delta\phi = \sqrt{\frac{1}{\mathrm{SNR}} + \frac{\gamma^2 S_\text{BB} \tau}{4 \pi^2}}.  \\
\label{eq:quadraturenoisesum}
\end{equation}
In the detector noise-limited regime, where $\delta\phi_\text{shot}^2 \gg \delta\phi_\text{field}^2$, this reduces to
\begin{equation}
\delta\phi = \sqrt{\frac{1}{\mathrm{SNR}}}.  \\
\label{eq:lowfieldnoiselimit}
\end{equation}
Combining this with Eq.~\eqref{eq:phasedeltasensitivity}, we find our governing intrinsic sensitivity equation
\begin{equation}
\delta B_\text{dc} = \frac{1}{\gamma \tau^{\frac{3}{2}}} \sqrt{\frac{12}{f_s \; \mathrm{SNR}}}. \\
\label{eq:finalsensitivity}
\end{equation}
It can additionally be shown that this achieves the Cramer-Rao bound for phase-based frequency estimation. The Cramer-Rao bound on the variance of the unbiased estimator of angular frequency is given~\cite{Kay3} by
\begin{equation}
\sigma^2_{\hat{\omega}} = \frac{12}{\mathrm{SNR} \; N (N^2 -1)} \; \left(\frac{\mathrm{rad}}{\mathrm{sample}}\right)^2,
\label{eq:cramerrao}
\end{equation} 
which in the large sample limit is
\begin{equation}
\sigma^2_{\hat{\omega}} = \frac{12 f_s^2}{\mathrm{SNR} \; N^3}.
\label{eq:highncramerrao}
\end{equation} 
Putting this in terms of previously defined quantities using $N = f_s \tau$ and $\omega = \gamma B$, this becomes
\begin{equation}
\sigma^2_{\hat{B}} = \frac{12} {\gamma^2 f_s \; \mathrm{SNR} \; \tau^3},
\label{eq:reframedhighncramerrao}
\end{equation}
\noindent or as a sensitivity,
\begin{equation}
\sigma_{\hat{B}} = \frac{1}{\gamma \tau^{\frac{3}{2}}} \sqrt{\frac{12}{f_s \mathrm{SNR}}}, \\
\label{eq:cramerraoderivedsensitivity}
\end{equation}
\noindent exactly equivalent to  Eq.~\eqref{eq:finalsensitivity}.

That least-squares regression in the high SNR limit achieves the Cramer-Rao bound for phase-based frequency estimation is an established result in information theory~\cite{Fowler}, shown here for completeness. In the case of a decaying SNR, as is the case for our Faraday measurement, the reconstructed phase estimates are heteroskedastic, and as such we estimate the Larmor frequency via a weighted least-squares regression, which is the best linear unbiased estimator~\cite{Aitken2}. Though the statistics of this regression are more involved, a weighted average SNR, which takes into account the decay lifetime and measurement duration, can be found, from which the detector-limited sensitivity can be calculated as per Eq.~\eqref{eq:finalsensitivity}.

\bibliography{apssamp}

\providecommand{\noopsort}[1]{}\providecommand{\singleletter}[1]{#1}%
\begin{thebibliography}{55}%
\makeatletter
\providecommand \@ifxundefined [1]{%
 \@ifx{#1\undefined}
}%
\providecommand \@ifnum [1]{%
 \ifnum #1\expandafter \@firstoftwo
 \else \expandafter \@secondoftwo
 \fi
}%
\providecommand \@ifx [1]{%
 \ifx #1\expandafter \@firstoftwo
 \else \expandafter \@secondoftwo
 \fi
}%
\providecommand \natexlab [1]{#1}%
\providecommand \enquote  [1]{``#1''}%
\providecommand \bibnamefont  [1]{#1}%
\providecommand \bibfnamefont [1]{#1}%
\providecommand \citenamefont [1]{#1}%
\providecommand \href@noop [0]{\@secondoftwo}%
\providecommand \href [0]{\begingroup \@sanitize@url \@href}%
\providecommand \@href[1]{\@@startlink{#1}\@@href}%
\providecommand \@@href[1]{\endgroup#1\@@endlink}%
\providecommand \@sanitize@url [0]{\catcode `\\12\catcode `\$12\catcode `\&12\catcode `\#12\catcode `\^12\catcode `\_12\catcode `\%12\relax}%
\providecommand \@@startlink[1]{}%
\providecommand \@@endlink[0]{}%
\providecommand \url  [0]{\begingroup\@sanitize@url \@url }%
\providecommand \@url [1]{\endgroup\@href {#1}{\urlprefix }}%
\providecommand \urlprefix  [0]{URL }%
\providecommand \Eprint [0]{\href }%
\providecommand \doibase [0]{https://doi.org/}%
\providecommand \selectlanguage [0]{\@gobble}%
\providecommand \bibinfo  [0]{\@secondoftwo}%
\providecommand \bibfield  [0]{\@secondoftwo}%
\providecommand \translation [1]{[#1]}%
\providecommand \BibitemOpen [0]{}%
\providecommand \bibitemStop [0]{}%
\providecommand \bibitemNoStop [0]{.\EOS\space}%
\providecommand \EOS [0]{\spacefactor3000\relax}%
\providecommand \BibitemShut  [1]{\csname bibitem#1\endcsname}%
\let\auto@bib@innerbib\@empty
\bibitem [{\citenamefont {Belfi}\ \emph {et~al.}(2007)\citenamefont {Belfi}, \citenamefont {Bevilacqua}, \citenamefont {Biancalana}, \citenamefont {Cartaleva}, \citenamefont {Dancheva},\ and\ \citenamefont {Moi}}]{Belfi}%
  \BibitemOpen
  \bibfield  {author} {\bibinfo {author} {\bibfnamefont {J.}~\bibnamefont {Belfi}}, \bibinfo {author} {\bibfnamefont {G.}~\bibnamefont {Bevilacqua}}, \bibinfo {author} {\bibfnamefont {V.}~\bibnamefont {Biancalana}}, \bibinfo {author} {\bibfnamefont {S.}~\bibnamefont {Cartaleva}}, \bibinfo {author} {\bibfnamefont {Y.}~\bibnamefont {Dancheva}},\ and\ \bibinfo {author} {\bibfnamefont {L.}~\bibnamefont {Moi}},\ }\bibfield  {title} {\bibinfo {title} {Cesium coherent population trapping magnetometer for cardiosignal detection in an unshielded environment},\ }\href {https://doi.org/10.1364/JOSAB.24.002357} {\bibfield  {journal} {\bibinfo  {journal} {J. Opt. Soc. Am. B}\ }\textbf {\bibinfo {volume} {24}},\ \bibinfo {pages} {2357} (\bibinfo {year} {2007})}\BibitemShut {NoStop}%
\bibitem [{\citenamefont {Xu}\ \emph {et~al.}(2006)\citenamefont {Xu}, \citenamefont {Yashchuk}, \citenamefont {Donaldson}, \citenamefont {Rochester}, \citenamefont {Budker},\ and\ \citenamefont {Pines}}]{Xu}%
  \BibitemOpen
  \bibfield  {author} {\bibinfo {author} {\bibfnamefont {S.}~\bibnamefont {Xu}}, \bibinfo {author} {\bibfnamefont {V.~V.}\ \bibnamefont {Yashchuk}}, \bibinfo {author} {\bibfnamefont {M.~H.}\ \bibnamefont {Donaldson}}, \bibinfo {author} {\bibfnamefont {S.~M.}\ \bibnamefont {Rochester}}, \bibinfo {author} {\bibfnamefont {D.}~\bibnamefont {Budker}},\ and\ \bibinfo {author} {\bibfnamefont {A.}~\bibnamefont {Pines}},\ }\bibfield  {title} {\bibinfo {title} {Magnetic resonance imaging with an optical atomic magnetometer},\ }\href {https://doi.org/10.1073/pnas.0605396103} {\bibfield  {journal} {\bibinfo  {journal} {Proceedings of the National Academy of Sciences}\ }\textbf {\bibinfo {volume} {103}},\ \bibinfo {pages} {12668} (\bibinfo {year} {2006})}\BibitemShut {NoStop}%
\bibitem [{\citenamefont {Wilson}\ \emph {et~al.}(2007)\citenamefont {Wilson}, \citenamefont {Tian},\ and\ \citenamefont {Barrans}}]{wilson2}%
  \BibitemOpen
  \bibfield  {author} {\bibinfo {author} {\bibfnamefont {J.~W.}\ \bibnamefont {Wilson}}, \bibinfo {author} {\bibfnamefont {G.~Y.}\ \bibnamefont {Tian}},\ and\ \bibinfo {author} {\bibfnamefont {S.}~\bibnamefont {Barrans}},\ }\bibfield  {title} {\bibinfo {title} {Residual magnetic field sensing for stress measurement},\ }\href {https://doi.org/https://doi.org/10.1016/j.sna.2006.08.010} {\bibfield  {journal} {\bibinfo  {journal} {Sensors and Actuators A}\ }\textbf {\bibinfo {volume} {135}},\ \bibinfo {pages} {381} (\bibinfo {year} {2007})}\BibitemShut {NoStop}%
\bibitem [{\citenamefont {Vengalattore}\ \emph {et~al.}(2007)\citenamefont {Vengalattore}, \citenamefont {Higbie}, \citenamefont {Leslie}, \citenamefont {Guzman}, \citenamefont {Sadler},\ and\ \citenamefont {Stamper-Kurn}}]{vengalattore}%
  \BibitemOpen
  \bibfield  {author} {\bibinfo {author} {\bibfnamefont {M.}~\bibnamefont {Vengalattore}}, \bibinfo {author} {\bibfnamefont {J.~M.}\ \bibnamefont {Higbie}}, \bibinfo {author} {\bibfnamefont {S.~R.}\ \bibnamefont {Leslie}}, \bibinfo {author} {\bibfnamefont {J.}~\bibnamefont {Guzman}}, \bibinfo {author} {\bibfnamefont {L.~E.}\ \bibnamefont {Sadler}},\ and\ \bibinfo {author} {\bibfnamefont {D.~M.}\ \bibnamefont {Stamper-Kurn}},\ }\bibfield  {title} {\bibinfo {title} {High-resolution magnetometry with a spinor bose-einstein condensate},\ }\href {https://doi.org/10.1103/PhysRevLett.98.200801} {\bibfield  {journal} {\bibinfo  {journal} {Physical Review Letters}\ }\textbf {\bibinfo {volume} {98}},\ \bibinfo {pages} {200801} (\bibinfo {year} {2007})}\BibitemShut {NoStop}%
\bibitem [{\citenamefont {Behbood}\ \emph {et~al.}(2013)\citenamefont {Behbood}, \citenamefont {Martin~Ciurana}, \citenamefont {Colangelo}, \citenamefont {Napolitano}, \citenamefont {Mitchell},\ and\ \citenamefont {Sewell}}]{Behbood}%
  \BibitemOpen
  \bibfield  {author} {\bibinfo {author} {\bibfnamefont {N.}~\bibnamefont {Behbood}}, \bibinfo {author} {\bibfnamefont {F.}~\bibnamefont {Martin~Ciurana}}, \bibinfo {author} {\bibfnamefont {G.}~\bibnamefont {Colangelo}}, \bibinfo {author} {\bibfnamefont {M.}~\bibnamefont {Napolitano}}, \bibinfo {author} {\bibfnamefont {M.~W.}\ \bibnamefont {Mitchell}},\ and\ \bibinfo {author} {\bibfnamefont {R.~J.}\ \bibnamefont {Sewell}},\ }\bibfield  {title} {\bibinfo {title} {Real-time vector field tracking with a cold-atom magnetometer},\ }\href {https://doi.org/10.1063/1.4803684} {\bibfield  {journal} {\bibinfo  {journal} {Applied Physics Letters}\ }\textbf {\bibinfo {volume} {102}},\ \bibinfo {pages} {173504} (\bibinfo {year} {2013})}\BibitemShut {NoStop}%
\bibitem [{\citenamefont {Yang}\ \emph {et~al.}(2017)\citenamefont {Yang}, \citenamefont {Koll\'ar}, \citenamefont {Taylor}, \citenamefont {Turner},\ and\ \citenamefont {Lev}}]{Yang}%
  \BibitemOpen
  \bibfield  {author} {\bibinfo {author} {\bibfnamefont {F.}~\bibnamefont {Yang}}, \bibinfo {author} {\bibfnamefont {A.~J.}\ \bibnamefont {Koll\'ar}}, \bibinfo {author} {\bibfnamefont {S.~F.}\ \bibnamefont {Taylor}}, \bibinfo {author} {\bibfnamefont {R.~W.}\ \bibnamefont {Turner}},\ and\ \bibinfo {author} {\bibfnamefont {B.~L.}\ \bibnamefont {Lev}},\ }\bibfield  {title} {\bibinfo {title} {Scanning quantum cryogenic atom microscope},\ }\href {https://doi.org/10.1103/PhysRevApplied.7.034026} {\bibfield  {journal} {\bibinfo  {journal} {Phys. Rev. Appl.}\ }\textbf {\bibinfo {volume} {7}},\ \bibinfo {pages} {034026} (\bibinfo {year} {2017})}\BibitemShut {NoStop}%
\bibitem [{\citenamefont {Eto}\ \emph {et~al.}(2013)\citenamefont {Eto}, \citenamefont {Ikeda}, \citenamefont {Suzuki}, \citenamefont {Hasegawa}, \citenamefont {Tomiyama}, \citenamefont {Sekine}, \citenamefont {Sadgrove},\ and\ \citenamefont {Hirano}}]{eto}%
  \BibitemOpen
  \bibfield  {author} {\bibinfo {author} {\bibfnamefont {Y.}~\bibnamefont {Eto}}, \bibinfo {author} {\bibfnamefont {H.}~\bibnamefont {Ikeda}}, \bibinfo {author} {\bibfnamefont {H.}~\bibnamefont {Suzuki}}, \bibinfo {author} {\bibfnamefont {S.}~\bibnamefont {Hasegawa}}, \bibinfo {author} {\bibfnamefont {Y.}~\bibnamefont {Tomiyama}}, \bibinfo {author} {\bibfnamefont {S.}~\bibnamefont {Sekine}}, \bibinfo {author} {\bibfnamefont {M.}~\bibnamefont {Sadgrove}},\ and\ \bibinfo {author} {\bibfnamefont {T.}~\bibnamefont {Hirano}},\ }\bibfield  {title} {\bibinfo {title} {Spin-echo-based magnetometry with spinor {Bose}-{Einstein} condensates},\ }\href {https://doi.org/10.1103/PhysRevA.88.031602} {\bibfield  {journal} {\bibinfo  {journal} {Physical Review A}\ }\textbf {\bibinfo {volume} {88}},\ \bibinfo {pages} {031602(R)} (\bibinfo {year} {2013})}\BibitemShut {NoStop}%
\bibitem [{\citenamefont {Cohen}\ \emph {et~al.}(2019)\citenamefont {Cohen}, \citenamefont {Jadeja}, \citenamefont {Sula}, \citenamefont {Venturelli}, \citenamefont {Deans}, \citenamefont {Marmugi},\ and\ \citenamefont {Renzoni}}]{cohen}%
  \BibitemOpen
  \bibfield  {author} {\bibinfo {author} {\bibfnamefont {Y.}~\bibnamefont {Cohen}}, \bibinfo {author} {\bibfnamefont {K.}~\bibnamefont {Jadeja}}, \bibinfo {author} {\bibfnamefont {S.}~\bibnamefont {Sula}}, \bibinfo {author} {\bibfnamefont {M.}~\bibnamefont {Venturelli}}, \bibinfo {author} {\bibfnamefont {C.}~\bibnamefont {Deans}}, \bibinfo {author} {\bibfnamefont {L.}~\bibnamefont {Marmugi}},\ and\ \bibinfo {author} {\bibfnamefont {F.}~\bibnamefont {Renzoni}},\ }\bibfield  {title} {\bibinfo {title} {A cold atom radio-frequency magnetometer},\ }\href {https://doi.org/10.1063/1.5084004} {\bibfield  {journal} {\bibinfo  {journal} {Applied Physics Letters}\ }\textbf {\bibinfo {volume} {114}},\ \bibinfo {pages} {073505} (\bibinfo {year} {2019})}\BibitemShut {NoStop}%
\bibitem [{\citenamefont {Mitchell}\ and\ \citenamefont {Palacios~Alvarez}(2020)}]{mitchell}%
  \BibitemOpen
  \bibfield  {author} {\bibinfo {author} {\bibfnamefont {M.~W.}\ \bibnamefont {Mitchell}}\ and\ \bibinfo {author} {\bibfnamefont {S.}~\bibnamefont {Palacios~Alvarez}},\ }\bibfield  {title} {\bibinfo {title} {Quantum limits to the energy resolution of magnetic field sensors},\ }\href {https://doi.org/10.1103/RevModPhys.92.021001} {\bibfield  {journal} {\bibinfo  {journal} {Reviews of Modern Physics}\ }\textbf {\bibinfo {volume} {92}},\ \bibinfo {pages} {021001} (\bibinfo {year} {2020})}\BibitemShut {NoStop}%
\bibitem [{\citenamefont {Chang}\ \emph {et~al.}(2017)\citenamefont {Chang}, \citenamefont {Eichler}, \citenamefont {Rhensius}, \citenamefont {Lorenzelli},\ and\ \citenamefont {Degen}}]{Chang}%
  \BibitemOpen
  \bibfield  {author} {\bibinfo {author} {\bibfnamefont {K.}~\bibnamefont {Chang}}, \bibinfo {author} {\bibfnamefont {A.}~\bibnamefont {Eichler}}, \bibinfo {author} {\bibfnamefont {J.}~\bibnamefont {Rhensius}}, \bibinfo {author} {\bibfnamefont {L.}~\bibnamefont {Lorenzelli}},\ and\ \bibinfo {author} {\bibfnamefont {C.~L.}\ \bibnamefont {Degen}},\ }\bibfield  {title} {\bibinfo {title} {Nanoscale imaging of current density with a single-spin magnetometer},\ }\href {https://doi.org/10.1021/acs.nanolett.6b05304} {\bibfield  {journal} {\bibinfo  {journal} {Nano Letters}\ }\textbf {\bibinfo {volume} {17}},\ \bibinfo {pages} {2367} (\bibinfo {year} {2017})},\ \bibinfo {note} {pMID: 28329445},\ \Eprint {https://arxiv.org/abs/https://doi.org/10.1021/acs.nanolett.6b05304} {https://doi.org/10.1021/acs.nanolett.6b05304} \BibitemShut {NoStop}%
\bibitem [{\citenamefont {Berg}\ \emph {et~al.}(2007)\citenamefont {Berg}, \citenamefont {Fradkin}, \citenamefont {Kim}, \citenamefont {Kivelson}, \citenamefont {Oganesyan}, \citenamefont {Tranquada},\ and\ \citenamefont {Zhang}}]{Berg}%
  \BibitemOpen
  \bibfield  {author} {\bibinfo {author} {\bibfnamefont {E.}~\bibnamefont {Berg}}, \bibinfo {author} {\bibfnamefont {E.}~\bibnamefont {Fradkin}}, \bibinfo {author} {\bibfnamefont {E.-A.}\ \bibnamefont {Kim}}, \bibinfo {author} {\bibfnamefont {S.~A.}\ \bibnamefont {Kivelson}}, \bibinfo {author} {\bibfnamefont {V.}~\bibnamefont {Oganesyan}}, \bibinfo {author} {\bibfnamefont {J.~M.}\ \bibnamefont {Tranquada}},\ and\ \bibinfo {author} {\bibfnamefont {S.~C.}\ \bibnamefont {Zhang}},\ }\bibfield  {title} {\bibinfo {title} {Dynamical layer decoupling in a stripe-ordered high-${T}_{c}$ superconductor},\ }\href {https://doi.org/10.1103/PhysRevLett.99.127003} {\bibfield  {journal} {\bibinfo  {journal} {Phys. Rev. Lett.}\ }\textbf {\bibinfo {volume} {99}},\ \bibinfo {pages} {127003} (\bibinfo {year} {2007})}\BibitemShut {NoStop}%
\bibitem [{\citenamefont {Griffith}\ \emph {et~al.}(2010)\citenamefont {Griffith}, \citenamefont {Knappe},\ and\ \citenamefont {Kitching}}]{griffith}%
  \BibitemOpen
  \bibfield  {author} {\bibinfo {author} {\bibfnamefont {W.~C.}\ \bibnamefont {Griffith}}, \bibinfo {author} {\bibfnamefont {S.}~\bibnamefont {Knappe}},\ and\ \bibinfo {author} {\bibfnamefont {J.}~\bibnamefont {Kitching}},\ }\bibfield  {title} {\bibinfo {title} {Femtotesla atomic magnetometry in a microfabricated vapor cell},\ }\href {https://doi.org/10.1364/OE.18.027167} {\bibfield  {journal} {\bibinfo  {journal} {Optics Express}\ }\textbf {\bibinfo {volume} {18}},\ \bibinfo {pages} {27167} (\bibinfo {year} {2010})}\BibitemShut {NoStop}%
\bibitem [{\citenamefont {Kominis}\ \emph {et~al.}(2003)\citenamefont {Kominis}, \citenamefont {Kornack}, \citenamefont {Allred},\ and\ \citenamefont {Romalis}}]{kominis}%
  \BibitemOpen
  \bibfield  {author} {\bibinfo {author} {\bibfnamefont {I.~K.}\ \bibnamefont {Kominis}}, \bibinfo {author} {\bibfnamefont {T.~W.}\ \bibnamefont {Kornack}}, \bibinfo {author} {\bibfnamefont {J.~C.}\ \bibnamefont {Allred}},\ and\ \bibinfo {author} {\bibfnamefont {M.~V.}\ \bibnamefont {Romalis}},\ }\bibfield  {title} {\bibinfo {title} {A subfemtotesla multichannel atomic magnetometer},\ }\href {https://doi.org/10.1038/nature01484} {\bibfield  {journal} {\bibinfo  {journal} {Nature}\ }\textbf {\bibinfo {volume} {422}},\ \bibinfo {pages} {596} (\bibinfo {year} {2003})}\BibitemShut {NoStop}%
\bibitem [{\citenamefont {Sheng}\ \emph {et~al.}(2013)\citenamefont {Sheng}, \citenamefont {Li}, \citenamefont {Dural},\ and\ \citenamefont {Romalis}}]{sheng}%
  \BibitemOpen
  \bibfield  {author} {\bibinfo {author} {\bibfnamefont {D.}~\bibnamefont {Sheng}}, \bibinfo {author} {\bibfnamefont {S.}~\bibnamefont {Li}}, \bibinfo {author} {\bibfnamefont {N.}~\bibnamefont {Dural}},\ and\ \bibinfo {author} {\bibfnamefont {M.~V.}\ \bibnamefont {Romalis}},\ }\bibfield  {title} {\bibinfo {title} {Subfemtotesla scalar atomic magnetometry using multipass cells},\ }\href {https://doi.org/10.1103/PhysRevLett.110.160802} {\bibfield  {journal} {\bibinfo  {journal} {Physical Review Letters}\ }\textbf {\bibinfo {volume} {110}},\ \bibinfo {pages} {160802} (\bibinfo {year} {2013})}\BibitemShut {NoStop}%
\bibitem [{\citenamefont {Dang}\ \emph {et~al.}(2010)\citenamefont {Dang}, \citenamefont {Maloof},\ and\ \citenamefont {Romalis}}]{dang}%
  \BibitemOpen
  \bibfield  {author} {\bibinfo {author} {\bibfnamefont {H.~B.}\ \bibnamefont {Dang}}, \bibinfo {author} {\bibfnamefont {A.~C.}\ \bibnamefont {Maloof}},\ and\ \bibinfo {author} {\bibfnamefont {M.~V.}\ \bibnamefont {Romalis}},\ }\bibfield  {title} {\bibinfo {title} {Ultrahigh sensitivity magnetic field and magnetization measurements with an atomic magnetometer},\ }\href {https://doi.org/10.1063/1.3491215} {\bibfield  {journal} {\bibinfo  {journal} {Applied Physics Letters}\ }\textbf {\bibinfo {volume} {97}},\ \bibinfo {pages} {151110} (\bibinfo {year} {2010})}\BibitemShut {NoStop}%
\bibitem [{\citenamefont {Bloom}(1962)}]{bloom}%
  \BibitemOpen
  \bibfield  {author} {\bibinfo {author} {\bibfnamefont {A.~L.}\ \bibnamefont {Bloom}},\ }\bibfield  {title} {\bibinfo {title} {Principles of operation of the rubidium vapor magnetometer},\ }\href {https://doi.org/10.1364/AO.1.000061} {\bibfield  {journal} {\bibinfo  {journal} {Appl. Opt.}\ }\textbf {\bibinfo {volume} {1}},\ \bibinfo {pages} {61} (\bibinfo {year} {1962})}\BibitemShut {NoStop}%
\bibitem [{\citenamefont {Chalupczak}\ \emph {et~al.}(2012)\citenamefont {Chalupczak}, \citenamefont {Godun}, \citenamefont {Pustelny},\ and\ \citenamefont {Gawlik}}]{chalupczak}%
  \BibitemOpen
  \bibfield  {author} {\bibinfo {author} {\bibfnamefont {W.}~\bibnamefont {Chalupczak}}, \bibinfo {author} {\bibfnamefont {R.~M.}\ \bibnamefont {Godun}}, \bibinfo {author} {\bibfnamefont {S.}~\bibnamefont {Pustelny}},\ and\ \bibinfo {author} {\bibfnamefont {W.}~\bibnamefont {Gawlik}},\ }\bibfield  {title} {\bibinfo {title} {Room temperature femtotesla radio-frequency atomic magnetometer},\ }\href {https://doi.org/10.1063/1.4729016} {\bibfield  {journal} {\bibinfo  {journal} {Applied Physics Letters}\ }\textbf {\bibinfo {volume} {100}},\ \bibinfo {pages} {242401} (\bibinfo {year} {2012})}\BibitemShut {NoStop}%
\bibitem [{\citenamefont {Budker}\ and\ \citenamefont {Romalis}(2007)}]{Budker}%
  \BibitemOpen
  \bibfield  {author} {\bibinfo {author} {\bibfnamefont {D.}~\bibnamefont {Budker}}\ and\ \bibinfo {author} {\bibfnamefont {M.~V.}\ \bibnamefont {Romalis}},\ }\bibfield  {title} {\bibinfo {title} {Optical magnetometry},\ }\href {https://doi.org/10.1038/nphys566} {\bibfield  {journal} {\bibinfo  {journal} {Nature Physics}\ }\textbf {\bibinfo {volume} {3}},\ \bibinfo {pages} {227} (\bibinfo {year} {2007})}\BibitemShut {NoStop}%
\bibitem [{\citenamefont {Borna}\ \emph {et~al.}(2019)\citenamefont {Borna}, \citenamefont {Carter}, \citenamefont {DeRego}, \citenamefont {James},\ and\ \citenamefont {Schwindt}}]{Borna}%
  \BibitemOpen
  \bibfield  {author} {\bibinfo {author} {\bibfnamefont {A.}~\bibnamefont {Borna}}, \bibinfo {author} {\bibfnamefont {T.~R.}\ \bibnamefont {Carter}}, \bibinfo {author} {\bibfnamefont {P.}~\bibnamefont {DeRego}}, \bibinfo {author} {\bibfnamefont {C.~D.}\ \bibnamefont {James}},\ and\ \bibinfo {author} {\bibfnamefont {P.~D.~D.}\ \bibnamefont {Schwindt}},\ }\bibfield  {title} {\bibinfo {title} {Magnetic source imaging using a pulsed optically pumped magnetometer array},\ }\href {https://doi.org/10.1109/TIM.2018.2851458} {\bibfield  {journal} {\bibinfo  {journal} {IEEE Transactions on Instrumentation and Measurement}\ }\textbf {\bibinfo {volume} {68}},\ \bibinfo {pages} {493} (\bibinfo {year} {2019})}\BibitemShut {NoStop}%
\bibitem [{\citenamefont {Hardman}\ \emph {et~al.}(2016)\citenamefont {Hardman}, \citenamefont {Everitt}, \citenamefont {McDonald}, \citenamefont {Manju}, \citenamefont {Wigley}, \citenamefont {Sooriyabandara}, \citenamefont {Kuhn}, \citenamefont {Debs}, \citenamefont {Close},\ and\ \citenamefont {Robins}}]{Hardman}%
  \BibitemOpen
  \bibfield  {author} {\bibinfo {author} {\bibfnamefont {K.~S.}\ \bibnamefont {Hardman}}, \bibinfo {author} {\bibfnamefont {P.~J.}\ \bibnamefont {Everitt}}, \bibinfo {author} {\bibfnamefont {G.~D.}\ \bibnamefont {McDonald}}, \bibinfo {author} {\bibfnamefont {P.}~\bibnamefont {Manju}}, \bibinfo {author} {\bibfnamefont {P.~B.}\ \bibnamefont {Wigley}}, \bibinfo {author} {\bibfnamefont {M.~A.}\ \bibnamefont {Sooriyabandara}}, \bibinfo {author} {\bibfnamefont {C.~C.~N.}\ \bibnamefont {Kuhn}}, \bibinfo {author} {\bibfnamefont {J.~E.}\ \bibnamefont {Debs}}, \bibinfo {author} {\bibfnamefont {J.~D.}\ \bibnamefont {Close}},\ and\ \bibinfo {author} {\bibfnamefont {N.~P.}\ \bibnamefont {Robins}},\ }\bibfield  {title} {\bibinfo {title} {Simultaneous precision gravimetry and magnetic gradiometry with a {Bose}-{Einstein} condensate: A high precision, quantum sensor},\ }\href {https://doi.org/10.1103/PhysRevLett.117.138501} {\bibfield  {journal} {\bibinfo  {journal} {Physical Review Letters}\ }\textbf {\bibinfo {volume}
  {117}},\ \bibinfo {pages} {138501} (\bibinfo {year} {2016})}\BibitemShut {NoStop}%
\bibitem [{\citenamefont {Degen}\ \emph {et~al.}(2017)\citenamefont {Degen}, \citenamefont {Reinhard},\ and\ \citenamefont {Cappellaro}}]{Degen}%
  \BibitemOpen
  \bibfield  {author} {\bibinfo {author} {\bibfnamefont {C.~L.}\ \bibnamefont {Degen}}, \bibinfo {author} {\bibfnamefont {F.}~\bibnamefont {Reinhard}},\ and\ \bibinfo {author} {\bibfnamefont {P.}~\bibnamefont {Cappellaro}},\ }\bibfield  {title} {\bibinfo {title} {Quantum sensing},\ }\href {https://doi.org/10.1103/RevModPhys.89.035002} {\bibfield  {journal} {\bibinfo  {journal} {Rev. Mod. Phys.}\ }\textbf {\bibinfo {volume} {89}},\ \bibinfo {pages} {035002} (\bibinfo {year} {2017})}\BibitemShut {NoStop}%
\bibitem [{\citenamefont {Shiga}\ and\ \citenamefont {Takeuchi}(2012)}]{Shiga}%
  \BibitemOpen
  \bibfield  {author} {\bibinfo {author} {\bibfnamefont {N.}~\bibnamefont {Shiga}}\ and\ \bibinfo {author} {\bibfnamefont {M.}~\bibnamefont {Takeuchi}},\ }\bibfield  {title} {\bibinfo {title} {Locking the local oscillator phase to the atomic phase via weak measurement},\ }\href {https://doi.org/10.1088/1367-2630/14/2/023034} {\bibfield  {journal} {\bibinfo  {journal} {New Journal of Physics}\ }\textbf {\bibinfo {volume} {14}},\ \bibinfo {pages} {023034} (\bibinfo {year} {2012})}\BibitemShut {NoStop}%
\bibitem [{\citenamefont {Lin}\ \emph {et~al.}(2009)\citenamefont {Lin}, \citenamefont {Perry}, \citenamefont {Compton}, \citenamefont {Spielman},\ and\ \citenamefont {Porto}}]{Lin}%
  \BibitemOpen
  \bibfield  {author} {\bibinfo {author} {\bibfnamefont {Y.-J.}\ \bibnamefont {Lin}}, \bibinfo {author} {\bibfnamefont {A.~R.}\ \bibnamefont {Perry}}, \bibinfo {author} {\bibfnamefont {R.~L.}\ \bibnamefont {Compton}}, \bibinfo {author} {\bibfnamefont {I.~B.}\ \bibnamefont {Spielman}},\ and\ \bibinfo {author} {\bibfnamefont {J.~V.}\ \bibnamefont {Porto}},\ }\bibfield  {title} {\bibinfo {title} {Rapid production of $^{87}\text{R}\text{b}$ bose-einstein condensates in a combined magnetic and optical potential},\ }\href {https://doi.org/10.1103/PhysRevA.79.063631} {\bibfield  {journal} {\bibinfo  {journal} {Phys. Rev. A}\ }\textbf {\bibinfo {volume} {79}},\ \bibinfo {pages} {063631} (\bibinfo {year} {2009})}\BibitemShut {NoStop}%
\bibitem [{\citenamefont {Wood}(2017)}]{Wood2}%
  \BibitemOpen
  \bibfield  {author} {\bibinfo {author} {\bibfnamefont {A.}~\bibnamefont {Wood}},\ }\emph {\bibinfo {title} {{Spinor Bose-Einstein condensates in magnetic field gradients}}},\ \href {https://doi.org/10.4225/03/58b3ba6e69f9a} {Ph.D. thesis},\ \bibinfo  {school} {Monash University} (\bibinfo {year} {2017})\BibitemShut {NoStop}%
\bibitem [{\citenamefont {Hong}\ \emph {et~al.}(2021)\citenamefont {Hong}, \citenamefont {Corrodi}, \citenamefont {Charity}, \citenamefont {Baeßler}, \citenamefont {Bono}, \citenamefont {Chupp}, \citenamefont {Fertl}, \citenamefont {Flay}, \citenamefont {García}, \citenamefont {George}, \citenamefont {Giovanetti}, \citenamefont {Gorringe}, \citenamefont {Grange}, \citenamefont {Hong}, \citenamefont {Kawall}, \citenamefont {Kiburg}, \citenamefont {Li}, \citenamefont {Li}, \citenamefont {Osofsky}, \citenamefont {Počanić}, \citenamefont {Ramachandran}, \citenamefont {Smith}, \citenamefont {Swanson}, \citenamefont {Tewsley-Booth}, \citenamefont {Winter}, \citenamefont {Yang},\ and\ \citenamefont {Zheng}}]{Hong}%
  \BibitemOpen
  \bibfield  {author} {\bibinfo {author} {\bibfnamefont {R.}~\bibnamefont {Hong}}, \bibinfo {author} {\bibfnamefont {S.}~\bibnamefont {Corrodi}}, \bibinfo {author} {\bibfnamefont {S.}~\bibnamefont {Charity}}, \bibinfo {author} {\bibfnamefont {S.}~\bibnamefont {Baeßler}}, \bibinfo {author} {\bibfnamefont {J.}~\bibnamefont {Bono}}, \bibinfo {author} {\bibfnamefont {T.}~\bibnamefont {Chupp}}, \bibinfo {author} {\bibfnamefont {M.}~\bibnamefont {Fertl}}, \bibinfo {author} {\bibfnamefont {D.}~\bibnamefont {Flay}}, \bibinfo {author} {\bibfnamefont {A.}~\bibnamefont {García}}, \bibinfo {author} {\bibfnamefont {J.}~\bibnamefont {George}}, \bibinfo {author} {\bibfnamefont {K.~L.}\ \bibnamefont {Giovanetti}}, \bibinfo {author} {\bibfnamefont {T.}~\bibnamefont {Gorringe}}, \bibinfo {author} {\bibfnamefont {J.}~\bibnamefont {Grange}}, \bibinfo {author} {\bibfnamefont {K.~W.}\ \bibnamefont {Hong}}, \bibinfo {author} {\bibfnamefont {D.}~\bibnamefont {Kawall}}, \bibinfo {author} {\bibfnamefont {B.}~\bibnamefont {Kiburg}},
  \bibinfo {author} {\bibfnamefont {B.}~\bibnamefont {Li}}, \bibinfo {author} {\bibfnamefont {L.}~\bibnamefont {Li}}, \bibinfo {author} {\bibfnamefont {R.}~\bibnamefont {Osofsky}}, \bibinfo {author} {\bibfnamefont {D.}~\bibnamefont {Počanić}}, \bibinfo {author} {\bibfnamefont {S.}~\bibnamefont {Ramachandran}}, \bibinfo {author} {\bibfnamefont {M.}~\bibnamefont {Smith}}, \bibinfo {author} {\bibfnamefont {H.~E.}\ \bibnamefont {Swanson}}, \bibinfo {author} {\bibfnamefont {A.}~\bibnamefont {Tewsley-Booth}}, \bibinfo {author} {\bibfnamefont {P.}~\bibnamefont {Winter}}, \bibinfo {author} {\bibfnamefont {T.}~\bibnamefont {Yang}},\ and\ \bibinfo {author} {\bibfnamefont {K.}~\bibnamefont {Zheng}},\ }\bibfield  {title} {\bibinfo {title} {Systematic and statistical uncertainties of the hilbert-transform based high-precision {FID} frequency extraction method},\ }\href {https://doi.org/10.1016/j.jmr.2021.107020} {\bibfield  {journal} {\bibinfo  {journal} {Journal of Magnetic Resonance}\ }\textbf {\bibinfo {volume}
  {329}},\ \bibinfo {pages} {107020} (\bibinfo {year} {2021})}\BibitemShut {NoStop}%
\bibitem [{\citenamefont {Wilson}\ \emph {et~al.}(2020)\citenamefont {Wilson}, \citenamefont {Perrella}, \citenamefont {Anderson}, \citenamefont {Luiten},\ and\ \citenamefont {Light}}]{Wilson}%
  \BibitemOpen
  \bibfield  {author} {\bibinfo {author} {\bibfnamefont {N.}~\bibnamefont {Wilson}}, \bibinfo {author} {\bibfnamefont {C.}~\bibnamefont {Perrella}}, \bibinfo {author} {\bibfnamefont {R.}~\bibnamefont {Anderson}}, \bibinfo {author} {\bibfnamefont {A.}~\bibnamefont {Luiten}},\ and\ \bibinfo {author} {\bibfnamefont {P.}~\bibnamefont {Light}},\ }\bibfield  {title} {\bibinfo {title} {Wide-bandwidth atomic magnetometry via instantaneous-phase retrieval},\ }\href {https://doi.org/10.1103/PhysRevResearch.2.013213} {\bibfield  {journal} {\bibinfo  {journal} {Physical Review Research}\ }\textbf {\bibinfo {volume} {2}},\ \bibinfo {pages} {013213} (\bibinfo {year} {2020})}\BibitemShut {NoStop}%
\bibitem [{\citenamefont {Blanes}\ \emph {et~al.}(2009)\citenamefont {Blanes}, \citenamefont {Casas}, \citenamefont {Oteo},\ and\ \citenamefont {Ros}}]{blanes}%
  \BibitemOpen
  \bibfield  {author} {\bibinfo {author} {\bibfnamefont {S.}~\bibnamefont {Blanes}}, \bibinfo {author} {\bibfnamefont {F.}~\bibnamefont {Casas}}, \bibinfo {author} {\bibfnamefont {J.~A.}\ \bibnamefont {Oteo}},\ and\ \bibinfo {author} {\bibfnamefont {J.}~\bibnamefont {Ros}},\ }\bibfield  {title} {\bibinfo {title} {The {Magnus} expansion and some of its applications},\ }\href {https://doi.org/10.1016/j.physrep.2008.11.001} {\bibfield  {journal} {\bibinfo  {journal} {Physics Reports}\ }\textbf {\bibinfo {volume} {470}},\ \bibinfo {pages} {151} (\bibinfo {year} {2009})}\BibitemShut {NoStop}%
\bibitem [{\citenamefont {Bounds}\ \emph {et~al.}(2024)\citenamefont {Bounds}, \citenamefont {Duff}, \citenamefont {Tritt}, \citenamefont {Taylor}, \citenamefont {Coe}, \citenamefont {White},\ and\ \citenamefont {Turner}}]{Bounds}%
  \BibitemOpen
  \bibfield  {author} {\bibinfo {author} {\bibfnamefont {C.~C.}\ \bibnamefont {Bounds}}, \bibinfo {author} {\bibfnamefont {J.~P.}\ \bibnamefont {Duff}}, \bibinfo {author} {\bibfnamefont {A.}~\bibnamefont {Tritt}}, \bibinfo {author} {\bibfnamefont {H.~A.~M.}\ \bibnamefont {Taylor}}, \bibinfo {author} {\bibfnamefont {G.~X.}\ \bibnamefont {Coe}}, \bibinfo {author} {\bibfnamefont {S.~J.}\ \bibnamefont {White}},\ and\ \bibinfo {author} {\bibfnamefont {L.~D.}\ \bibnamefont {Turner}},\ }\bibfield  {title} {\bibinfo {title} {Quantum spectral analysis by continuous measurement of landau-zener transitions},\ }\href {https://doi.org/10.1103/PhysRevLett.132.093401} {\bibfield  {journal} {\bibinfo  {journal} {Phys. Rev. Lett.}\ }\textbf {\bibinfo {volume} {132}},\ \bibinfo {pages} {093401} (\bibinfo {year} {2024})}\BibitemShut {NoStop}%
\bibitem [{\citenamefont {Taylor}\ \emph {et~al.}(2008)\citenamefont {Taylor}, \citenamefont {Cappellaro}, \citenamefont {Childress}, \citenamefont {Jiang}, \citenamefont {Budker}, \citenamefont {Hemmer}, \citenamefont {Yacoby}, \citenamefont {Walsworth},\ and\ \citenamefont {Lukin}}]{taylor}%
  \BibitemOpen
  \bibfield  {author} {\bibinfo {author} {\bibfnamefont {J.~M.}\ \bibnamefont {Taylor}}, \bibinfo {author} {\bibfnamefont {P.}~\bibnamefont {Cappellaro}}, \bibinfo {author} {\bibfnamefont {L.}~\bibnamefont {Childress}}, \bibinfo {author} {\bibfnamefont {L.}~\bibnamefont {Jiang}}, \bibinfo {author} {\bibfnamefont {D.}~\bibnamefont {Budker}}, \bibinfo {author} {\bibfnamefont {P.~R.}\ \bibnamefont {Hemmer}}, \bibinfo {author} {\bibfnamefont {A.}~\bibnamefont {Yacoby}}, \bibinfo {author} {\bibfnamefont {R.}~\bibnamefont {Walsworth}},\ and\ \bibinfo {author} {\bibfnamefont {M.~D.}\ \bibnamefont {Lukin}},\ }\bibfield  {title} {\bibinfo {title} {High-sensitivity diamond magnetometer with nanoscale resolution},\ }\href {https://doi.org/10.1038/nphys1075} {\bibfield  {journal} {\bibinfo  {journal} {Nature Physics}\ }\textbf {\bibinfo {volume} {4}},\ \bibinfo {pages} {810} (\bibinfo {year} {2008})}\BibitemShut {NoStop}%
\bibitem [{\citenamefont {Waldherr}\ \emph {et~al.}(2011)\citenamefont {Waldherr}, \citenamefont {Beck}, \citenamefont {Neumann}, \citenamefont {Said}, \citenamefont {Nitsche}, \citenamefont {Markham}, \citenamefont {Twitchen}, \citenamefont {Twamley},\ and\ \citenamefont {Jelezko}}]{Waldherr}%
  \BibitemOpen
  \bibfield  {author} {\bibinfo {author} {\bibfnamefont {G.}~\bibnamefont {Waldherr}}, \bibinfo {author} {\bibfnamefont {J.}~\bibnamefont {Beck}}, \bibinfo {author} {\bibfnamefont {P.}~\bibnamefont {Neumann}}, \bibinfo {author} {\bibfnamefont {R.}~\bibnamefont {Said}}, \bibinfo {author} {\bibfnamefont {M.}~\bibnamefont {Nitsche}}, \bibinfo {author} {\bibfnamefont {M.}~\bibnamefont {Markham}}, \bibinfo {author} {\bibfnamefont {D.}~\bibnamefont {Twitchen}}, \bibinfo {author} {\bibfnamefont {J.}~\bibnamefont {Twamley}},\ and\ \bibinfo {author} {\bibfnamefont {F.}~\bibnamefont {Jelezko}},\ }\bibfield  {title} {\bibinfo {title} {High-dynamic-range magnetometry with a single nuclear spin in diamond},\ }\href {https://doi.org/10.1038/nnano.2011.224} {\bibfield  {journal} {\bibinfo  {journal} {Nature nanotechnology}\ }\textbf {\bibinfo {volume} {7}},\ \bibinfo {pages} {105} (\bibinfo {year} {2011})}\BibitemShut {NoStop}%
\bibitem [{\citenamefont {Wöltgens}\ and\ \citenamefont {Koch}(2000)}]{Woltgens}%
  \BibitemOpen
  \bibfield  {author} {\bibinfo {author} {\bibfnamefont {P.~J.~M.}\ \bibnamefont {Wöltgens}}\ and\ \bibinfo {author} {\bibfnamefont {R.~H.}\ \bibnamefont {Koch}},\ }\bibfield  {title} {\bibinfo {title} {Magnetic background noise cancellation in real-world environments},\ }\href {https://doi.org/10.1063/1.1150490} {\bibfield  {journal} {\bibinfo  {journal} {Review of Scientific Instruments}\ }\textbf {\bibinfo {volume} {71}},\ \bibinfo {pages} {1529} (\bibinfo {year} {2000})}\BibitemShut {NoStop}%
\bibitem [{\citenamefont {Deutsch}\ \emph {et~al.}(2010)\citenamefont {Deutsch}, \citenamefont {Ramirez-Martinez}, \citenamefont {Lacro\^ute}, \citenamefont {Reinhard}, \citenamefont {Schneider}, \citenamefont {Fuchs}, \citenamefont {Pi\'echon}, \citenamefont {Lalo\"e}, \citenamefont {Reichel},\ and\ \citenamefont {Rosenbusch}}]{Deutsch}%
  \BibitemOpen
  \bibfield  {author} {\bibinfo {author} {\bibfnamefont {C.}~\bibnamefont {Deutsch}}, \bibinfo {author} {\bibfnamefont {F.}~\bibnamefont {Ramirez-Martinez}}, \bibinfo {author} {\bibfnamefont {C.}~\bibnamefont {Lacro\^ute}}, \bibinfo {author} {\bibfnamefont {F.}~\bibnamefont {Reinhard}}, \bibinfo {author} {\bibfnamefont {T.}~\bibnamefont {Schneider}}, \bibinfo {author} {\bibfnamefont {J.~N.}\ \bibnamefont {Fuchs}}, \bibinfo {author} {\bibfnamefont {F.}~\bibnamefont {Pi\'echon}}, \bibinfo {author} {\bibfnamefont {F.}~\bibnamefont {Lalo\"e}}, \bibinfo {author} {\bibfnamefont {J.}~\bibnamefont {Reichel}},\ and\ \bibinfo {author} {\bibfnamefont {P.}~\bibnamefont {Rosenbusch}},\ }\bibfield  {title} {\bibinfo {title} {Spin self-rephasing and very long coherence times in a trapped atomic ensemble},\ }\href {https://doi.org/10.1103/PhysRevLett.105.020401} {\bibfield  {journal} {\bibinfo  {journal} {Physical Review Letters}\ }\textbf {\bibinfo {volume} {105}},\ \bibinfo {pages} {020401} (\bibinfo {year}
  {2010})}\BibitemShut {NoStop}%
\bibitem [{\citenamefont {Jasperse}\ \emph {et~al.}(2017)\citenamefont {Jasperse}, \citenamefont {Kewming}, \citenamefont {Fischer}, \citenamefont {Pakkiam}, \citenamefont {Anderson},\ and\ \citenamefont {Turner}}]{Jasperse}%
  \BibitemOpen
  \bibfield  {author} {\bibinfo {author} {\bibfnamefont {M.}~\bibnamefont {Jasperse}}, \bibinfo {author} {\bibfnamefont {M.~J.}\ \bibnamefont {Kewming}}, \bibinfo {author} {\bibfnamefont {S.~N.}\ \bibnamefont {Fischer}}, \bibinfo {author} {\bibfnamefont {P.}~\bibnamefont {Pakkiam}}, \bibinfo {author} {\bibfnamefont {R.~P.}\ \bibnamefont {Anderson}},\ and\ \bibinfo {author} {\bibfnamefont {L.~D.}\ \bibnamefont {Turner}},\ }\bibfield  {title} {\bibinfo {title} {Continuous {Faraday} measurement of spin precession without light shifts},\ }\href {https://doi.org/10.1103/PhysRevA.96.063402} {\bibfield  {journal} {\bibinfo  {journal} {Physical Review A}\ }\textbf {\bibinfo {volume} {96}},\ \bibinfo {pages} {063402} (\bibinfo {year} {2017})}\BibitemShut {NoStop}%
\bibitem [{\citenamefont {Jensen}\ \emph {et~al.}(2009)\citenamefont {Jensen}, \citenamefont {Acosta}, \citenamefont {Higbie}, \citenamefont {Ledbetter}, \citenamefont {Rochester},\ and\ \citenamefont {Budker}}]{Jensen}%
  \BibitemOpen
  \bibfield  {author} {\bibinfo {author} {\bibfnamefont {K.}~\bibnamefont {Jensen}}, \bibinfo {author} {\bibfnamefont {V.~M.}\ \bibnamefont {Acosta}}, \bibinfo {author} {\bibfnamefont {J.~M.}\ \bibnamefont {Higbie}}, \bibinfo {author} {\bibfnamefont {M.~P.}\ \bibnamefont {Ledbetter}}, \bibinfo {author} {\bibfnamefont {S.~M.}\ \bibnamefont {Rochester}},\ and\ \bibinfo {author} {\bibfnamefont {D.}~\bibnamefont {Budker}},\ }\bibfield  {title} {\bibinfo {title} {Cancellation of nonlinear zeeman shifts with light shifts},\ }\href {https://doi.org/10.1103/PhysRevA.79.023406} {\bibfield  {journal} {\bibinfo  {journal} {Phys. Rev. A}\ }\textbf {\bibinfo {volume} {79}},\ \bibinfo {pages} {023406} (\bibinfo {year} {2009})}\BibitemShut {NoStop}%
\bibitem [{\citenamefont {Chalupczak}\ \emph {et~al.}(2010)\citenamefont {Chalupczak}, \citenamefont {Wojciechowski}, \citenamefont {Pustelny},\ and\ \citenamefont {Gawlik}}]{Chalupczak2}%
  \BibitemOpen
  \bibfield  {author} {\bibinfo {author} {\bibfnamefont {W.}~\bibnamefont {Chalupczak}}, \bibinfo {author} {\bibfnamefont {A.}~\bibnamefont {Wojciechowski}}, \bibinfo {author} {\bibfnamefont {S.}~\bibnamefont {Pustelny}},\ and\ \bibinfo {author} {\bibfnamefont {W.}~\bibnamefont {Gawlik}},\ }\bibfield  {title} {\bibinfo {title} {Competition between the tensor light shift and nonlinear zeeman effect},\ }\href {https://doi.org/10.1103/PhysRevA.82.023417} {\bibfield  {journal} {\bibinfo  {journal} {Phys. Rev. A}\ }\textbf {\bibinfo {volume} {82}},\ \bibinfo {pages} {023417} (\bibinfo {year} {2010})}\BibitemShut {NoStop}%
\bibitem [{\citenamefont {Fowler}(2002)}]{Fowler}%
  \BibitemOpen
  \bibfield  {author} {\bibinfo {author} {\bibfnamefont {M.~L.}\ \bibnamefont {Fowler}},\ }\bibfield  {title} {\bibinfo {title} {Phase-based frequency estimation: A review},\ }\href {https://doi.org/10.1006/dspr.2001.0415} {\bibfield  {journal} {\bibinfo  {journal} {Digital Signal Processing}\ }\textbf {\bibinfo {volume} {12}},\ \bibinfo {pages} {590} (\bibinfo {year} {2002})}\BibitemShut {NoStop}%
\bibitem [{\citenamefont {Smith}\ \emph {et~al.}(2003)\citenamefont {Smith}, \citenamefont {Chaudhury},\ and\ \citenamefont {Jessen}}]{smith}%
  \BibitemOpen
  \bibfield  {author} {\bibinfo {author} {\bibfnamefont {G.~A.}\ \bibnamefont {Smith}}, \bibinfo {author} {\bibfnamefont {S.}~\bibnamefont {Chaudhury}},\ and\ \bibinfo {author} {\bibfnamefont {P.~S.}\ \bibnamefont {Jessen}},\ }\bibfield  {title} {\bibinfo {title} {Faraday spectroscopy in an optical lattice: a continuous probe of atom dynamics},\ }\href {https://doi.org/10.1088/1464-4266/5/4/301} {\bibfield  {journal} {\bibinfo  {journal} {J. Opt. Soc. Am. B}\ }\textbf {\bibinfo {volume} {5}},\ \bibinfo {pages} {323} (\bibinfo {year} {2003})}\BibitemShut {NoStop}%
\bibitem [{\citenamefont {Kim}\ \emph {et~al.}(1996)\citenamefont {Kim}, \citenamefont {Narasimha},\ and\ \citenamefont {Cox}}]{Kim}%
  \BibitemOpen
  \bibfield  {author} {\bibinfo {author} {\bibfnamefont {D.}~\bibnamefont {Kim}}, \bibinfo {author} {\bibfnamefont {M.}~\bibnamefont {Narasimha}},\ and\ \bibinfo {author} {\bibfnamefont {D.}~\bibnamefont {Cox}},\ }\bibfield  {title} {\bibinfo {title} {An improved single frequency estimator},\ }\href {https://doi.org/10.1109/97.508168} {\bibfield  {journal} {\bibinfo  {journal} {IEEE Signal Processing Letters}\ }\textbf {\bibinfo {volume} {3}},\ \bibinfo {pages} {212} (\bibinfo {year} {1996})}\BibitemShut {NoStop}%
\bibitem [{\citenamefont {Luise}\ and\ \citenamefont {Reggiannini}(1995)}]{Luise}%
  \BibitemOpen
  \bibfield  {author} {\bibinfo {author} {\bibfnamefont {M.}~\bibnamefont {Luise}}\ and\ \bibinfo {author} {\bibfnamefont {R.}~\bibnamefont {Reggiannini}},\ }\bibfield  {title} {\bibinfo {title} {Carrier frequency recovery in all-digital modems for burst-mode transmissions},\ }\href {https://doi.org/10.1109/26.380149} {\bibfield  {journal} {\bibinfo  {journal} {IEEE Transactions on Communications}\ }\textbf {\bibinfo {volume} {43}},\ \bibinfo {pages} {1169} (\bibinfo {year} {1995})}\BibitemShut {NoStop}%
\bibitem [{\citenamefont {Fitz}(1994)}]{Fitz}%
  \BibitemOpen
  \bibfield  {author} {\bibinfo {author} {\bibfnamefont {M.}~\bibnamefont {Fitz}},\ }\bibfield  {title} {\bibinfo {title} {Further results in the fast estimation of a single frequency},\ }\href {https://doi.org/10.1109/TCOMM.1994.580190} {\bibfield  {journal} {\bibinfo  {journal} {IEEE Transactions on Communications}\ }\textbf {\bibinfo {volume} {42}},\ \bibinfo {pages} {862} (\bibinfo {year} {1994})}\BibitemShut {NoStop}%
\bibitem [{\citenamefont {Zhao}\ \emph {et~al.}(2023)\citenamefont {Zhao}, \citenamefont {Särkkä}, \citenamefont {Sjölund},\ and\ \citenamefont {Schön}}]{zhao}%
  \BibitemOpen
  \bibfield  {author} {\bibinfo {author} {\bibfnamefont {Z.}~\bibnamefont {Zhao}}, \bibinfo {author} {\bibfnamefont {S.}~\bibnamefont {Särkkä}}, \bibinfo {author} {\bibfnamefont {J.}~\bibnamefont {Sjölund}},\ and\ \bibinfo {author} {\bibfnamefont {T.~B.}\ \bibnamefont {Schön}},\ }\bibfield  {title} {\bibinfo {title} {Probabilistic estimation of instantaneous frequencies of chirp signals},\ }\href {https://doi.org/10.1109/TSP.2023.3245720} {\bibfield  {journal} {\bibinfo  {journal} {IEEE Transactions on Signal Processing}\ }\textbf {\bibinfo {volume} {71}},\ \bibinfo {pages} {461} (\bibinfo {year} {2023})}\BibitemShut {NoStop}%
\bibitem [{\citenamefont {Bracewell}(1999)}]{bracewell}%
  \BibitemOpen
  \bibfield  {author} {\bibinfo {author} {\bibfnamefont {R.~N.}\ \bibnamefont {Bracewell}},\ }\href@noop {} {\emph {\bibinfo {title} {Fourier Transform and its Applications}}}\ (\bibinfo  {publisher} {McGraw-Hill},\ \bibinfo {year} {1999})\BibitemShut {NoStop}%
\bibitem [{\citenamefont {Harris}\ \emph {et~al.}(2020)\citenamefont {Harris}, \citenamefont {Millman}, \citenamefont {van~der Walt}, \citenamefont {Gommers}, \citenamefont {Virtanen}, \citenamefont {Cournapeau}, \citenamefont {Wieser}, \citenamefont {Taylor}, \citenamefont {Berg}, \citenamefont {Smith}, \citenamefont {Kern}, \citenamefont {Picus}, \citenamefont {Hoyer}, \citenamefont {van Kerkwijk}, \citenamefont {Brett}, \citenamefont {Haldane}, \citenamefont {del R{\'{i}}o}, \citenamefont {Wiebe}, \citenamefont {Peterson}, \citenamefont {G{\'{e}}rard-Marchant}, \citenamefont {Sheppard}, \citenamefont {Reddy}, \citenamefont {Weckesser}, \citenamefont {Abbasi}, \citenamefont {Gohlke},\ and\ \citenamefont {Oliphant}}]{harris}%
  \BibitemOpen
  \bibfield  {author} {\bibinfo {author} {\bibfnamefont {C.~R.}\ \bibnamefont {Harris}}, \bibinfo {author} {\bibfnamefont {K.~J.}\ \bibnamefont {Millman}}, \bibinfo {author} {\bibfnamefont {S.~J.}\ \bibnamefont {van~der Walt}}, \bibinfo {author} {\bibfnamefont {R.}~\bibnamefont {Gommers}}, \bibinfo {author} {\bibfnamefont {P.}~\bibnamefont {Virtanen}}, \bibinfo {author} {\bibfnamefont {D.}~\bibnamefont {Cournapeau}}, \bibinfo {author} {\bibfnamefont {E.}~\bibnamefont {Wieser}}, \bibinfo {author} {\bibfnamefont {J.}~\bibnamefont {Taylor}}, \bibinfo {author} {\bibfnamefont {S.}~\bibnamefont {Berg}}, \bibinfo {author} {\bibfnamefont {N.~J.}\ \bibnamefont {Smith}}, \bibinfo {author} {\bibfnamefont {R.}~\bibnamefont {Kern}}, \bibinfo {author} {\bibfnamefont {M.}~\bibnamefont {Picus}}, \bibinfo {author} {\bibfnamefont {S.}~\bibnamefont {Hoyer}}, \bibinfo {author} {\bibfnamefont {M.~H.}\ \bibnamefont {van Kerkwijk}}, \bibinfo {author} {\bibfnamefont {M.}~\bibnamefont {Brett}}, \bibinfo {author} {\bibfnamefont
  {A.}~\bibnamefont {Haldane}}, \bibinfo {author} {\bibfnamefont {J.~F.}\ \bibnamefont {del R{\'{i}}o}}, \bibinfo {author} {\bibfnamefont {M.}~\bibnamefont {Wiebe}}, \bibinfo {author} {\bibfnamefont {P.}~\bibnamefont {Peterson}}, \bibinfo {author} {\bibfnamefont {P.}~\bibnamefont {G{\'{e}}rard-Marchant}}, \bibinfo {author} {\bibfnamefont {K.}~\bibnamefont {Sheppard}}, \bibinfo {author} {\bibfnamefont {T.}~\bibnamefont {Reddy}}, \bibinfo {author} {\bibfnamefont {W.}~\bibnamefont {Weckesser}}, \bibinfo {author} {\bibfnamefont {H.}~\bibnamefont {Abbasi}}, \bibinfo {author} {\bibfnamefont {C.}~\bibnamefont {Gohlke}},\ and\ \bibinfo {author} {\bibfnamefont {T.~E.}\ \bibnamefont {Oliphant}},\ }\bibfield  {title} {\bibinfo {title} {Array programming with {NumPy}},\ }\href {https://doi.org/10.1038/s41586-020-2649-2} {\bibfield  {journal} {\bibinfo  {journal} {Nature}\ }\textbf {\bibinfo {volume} {585}},\ \bibinfo {pages} {357} (\bibinfo {year} {2020})}\BibitemShut {NoStop}%
\bibitem [{\citenamefont {Anderson}\ \emph {et~al.}(2018)\citenamefont {Anderson}, \citenamefont {Kewming},\ and\ \citenamefont {Turner}}]{anderson}%
  \BibitemOpen
  \bibfield  {author} {\bibinfo {author} {\bibfnamefont {R.~P.}\ \bibnamefont {Anderson}}, \bibinfo {author} {\bibfnamefont {M.~J.}\ \bibnamefont {Kewming}},\ and\ \bibinfo {author} {\bibfnamefont {L.~D.}\ \bibnamefont {Turner}},\ }\bibfield  {title} {\bibinfo {title} {Continuously observing a dynamically decoupled spin-1 quantum gas},\ }\href {https://doi.org/10.1103/PhysRevA.97.013408} {\bibfield  {journal} {\bibinfo  {journal} {Physical Review A}\ }\textbf {\bibinfo {volume} {97}},\ \bibinfo {pages} {013408} (\bibinfo {year} {2018})}\BibitemShut {NoStop}%
\bibitem [{\citenamefont {Carson}(1963)}]{Carson}%
  \BibitemOpen
  \bibfield  {author} {\bibinfo {author} {\bibfnamefont {J.}~\bibnamefont {Carson}},\ }\bibfield  {title} {\bibinfo {title} {Notes on the theory of modulation},\ }\href {https://doi.org/10.1109/PROC.1963.2322} {\bibfield  {journal} {\bibinfo  {journal} {Proceedings of the IEEE}\ }\textbf {\bibinfo {volume} {51}},\ \bibinfo {pages} {893} (\bibinfo {year} {1963})}\BibitemShut {NoStop}%
\bibitem [{\citenamefont {Merkel}\ \emph {et~al.}(2019)\citenamefont {Merkel}, \citenamefont {Thirumalai}, \citenamefont {Tarlton}, \citenamefont {Schäfer}, \citenamefont {Ballance}, \citenamefont {Harty},\ and\ \citenamefont {Lucas}}]{merkel}%
  \BibitemOpen
  \bibfield  {author} {\bibinfo {author} {\bibfnamefont {B.}~\bibnamefont {Merkel}}, \bibinfo {author} {\bibfnamefont {K.}~\bibnamefont {Thirumalai}}, \bibinfo {author} {\bibfnamefont {J.~E.}\ \bibnamefont {Tarlton}}, \bibinfo {author} {\bibfnamefont {V.~M.}\ \bibnamefont {Schäfer}}, \bibinfo {author} {\bibfnamefont {C.~J.}\ \bibnamefont {Ballance}}, \bibinfo {author} {\bibfnamefont {T.~P.}\ \bibnamefont {Harty}},\ and\ \bibinfo {author} {\bibfnamefont {D.~M.}\ \bibnamefont {Lucas}},\ }\bibfield  {title} {\bibinfo {title} {Magnetic field stabilization system for atomic physics experiments},\ }\href {https://doi.org/10.1063/1.5080093} {\bibfield  {journal} {\bibinfo  {journal} {Review of Scientific Instruments}\ }\textbf {\bibinfo {volume} {90}},\ \bibinfo {pages} {044702} (\bibinfo {year} {2019})}\BibitemShut {NoStop}%
\bibitem [{\citenamefont {{Platzek}}\ \emph {et~al.}(1999)\citenamefont {{Platzek}}, \citenamefont {{Nowak}}, \citenamefont {{Giessler}}, \citenamefont {{R{\"o}ther}},\ and\ \citenamefont {{Eiselt}}}]{Platzek}%
  \BibitemOpen
  \bibfield  {author} {\bibinfo {author} {\bibfnamefont {D.}~\bibnamefont {{Platzek}}}, \bibinfo {author} {\bibfnamefont {H.}~\bibnamefont {{Nowak}}}, \bibinfo {author} {\bibfnamefont {F.}~\bibnamefont {{Giessler}}}, \bibinfo {author} {\bibfnamefont {J.}~\bibnamefont {{R{\"o}ther}}},\ and\ \bibinfo {author} {\bibfnamefont {M.}~\bibnamefont {{Eiselt}}},\ }\bibfield  {title} {\bibinfo {title} {{Active shielding to reduce low frequency disturbances in direct current near biomagnetic measurements}},\ }\href {https://doi.org/10.1063/1.1149779} {\bibfield  {journal} {\bibinfo  {journal} {Review of Scientific Instruments}\ }\textbf {\bibinfo {volume} {70}},\ \bibinfo {pages} {2465} (\bibinfo {year} {1999})}\BibitemShut {NoStop}%
\bibitem [{\citenamefont {Xu}\ \emph {et~al.}(2019)\citenamefont {Xu}, \citenamefont {Wang}, \citenamefont {Jiao}, \citenamefont {Yi}, \citenamefont {Sun},\ and\ \citenamefont {Chen}}]{xu2}%
  \BibitemOpen
  \bibfield  {author} {\bibinfo {author} {\bibfnamefont {X.-T.}\ \bibnamefont {Xu}}, \bibinfo {author} {\bibfnamefont {Z.-Y.}\ \bibnamefont {Wang}}, \bibinfo {author} {\bibfnamefont {R.-H.}\ \bibnamefont {Jiao}}, \bibinfo {author} {\bibfnamefont {C.-R.}\ \bibnamefont {Yi}}, \bibinfo {author} {\bibfnamefont {W.}~\bibnamefont {Sun}},\ and\ \bibinfo {author} {\bibfnamefont {S.}~\bibnamefont {Chen}},\ }\bibfield  {title} {\bibinfo {title} {Ultra-low noise magnetic field for quantum gases},\ }\href {https://doi.org/10.1063/1.5087957} {\bibfield  {journal} {\bibinfo  {journal} {Review of Scientific Instruments}\ }\textbf {\bibinfo {volume} {90}},\ \bibinfo {pages} {054708} (\bibinfo {year} {2019})}\BibitemShut {NoStop}%
\bibitem [{\citenamefont {Zhang}\ \emph {et~al.}(2020)\citenamefont {Zhang}, \citenamefont {Ding}, \citenamefont {Yang}, \citenamefont {Zheng}, \citenamefont {Chen}, \citenamefont {Peng}, \citenamefont {Teng},\ and\ \citenamefont {Guo}}]{Zhang}%
  \BibitemOpen
  \bibfield  {author} {\bibinfo {author} {\bibfnamefont {R.}~\bibnamefont {Zhang}}, \bibinfo {author} {\bibfnamefont {Y.}~\bibnamefont {Ding}}, \bibinfo {author} {\bibfnamefont {Y.}~\bibnamefont {Yang}}, \bibinfo {author} {\bibfnamefont {Z.}~\bibnamefont {Zheng}}, \bibinfo {author} {\bibfnamefont {J.}~\bibnamefont {Chen}}, \bibinfo {author} {\bibfnamefont {X.}~\bibnamefont {Peng}}, \bibinfo {author} {\bibfnamefont {W.}~\bibnamefont {Teng}},\ and\ \bibinfo {author} {\bibfnamefont {H.}~\bibnamefont {Guo}},\ }\bibfield  {title} {\bibinfo {title} {Active magnetic-field stabilization with atomic magnetometer},\ }\href {https://doi.org/10.3390/s20154241} {\bibfield  {journal} {\bibinfo  {journal} {Sensors}\ }\textbf {\bibinfo {volume} {20}},\ \bibinfo {pages} {4241} (\bibinfo {year} {2020})}\BibitemShut {NoStop}%
\bibitem [{\citenamefont {Smith}\ \emph {et~al.}(2011)\citenamefont {Smith}, \citenamefont {Anderson}, \citenamefont {Chaudhury},\ and\ \citenamefont {Jessen}}]{smith3}%
  \BibitemOpen
  \bibfield  {author} {\bibinfo {author} {\bibfnamefont {A.}~\bibnamefont {Smith}}, \bibinfo {author} {\bibfnamefont {B.~E.}\ \bibnamefont {Anderson}}, \bibinfo {author} {\bibfnamefont {S.}~\bibnamefont {Chaudhury}},\ and\ \bibinfo {author} {\bibfnamefont {P.~S.}\ \bibnamefont {Jessen}},\ }\bibfield  {title} {\bibinfo {title} {Three-axis measurement and cancellation of background magnetic fields to less than 50 µ{G} in a cold atom experiment},\ }\href {https://doi.org/10.1088/0953-4075/44/20/205002} {\bibfield  {journal} {\bibinfo  {journal} {Journal of Physics B: Atomic, Molecular and Optical Physics}\ }\textbf {\bibinfo {volume} {44}},\ \bibinfo {pages} {205002} (\bibinfo {year} {2011})}\BibitemShut {NoStop}%
\bibitem [{\citenamefont {Panel}(2023)}]{aemc}%
  \BibitemOpen
  \bibfield  {author} {\bibinfo {author} {\bibfnamefont {R.}~\bibnamefont {Panel}},\ }\href@noop {} {\emph {\bibinfo {title} {The Frequency operating standard}}},\ \bibinfo {type} {Tech. Rep.}\ (\bibinfo  {institution} {Australian Energy Market Commission},\ \bibinfo {year} {2023})\BibitemShut {NoStop}%
\bibitem [{\citenamefont {Breit}\ and\ \citenamefont {Rabi}(1931)}]{breit}%
  \BibitemOpen
  \bibfield  {author} {\bibinfo {author} {\bibfnamefont {G.}~\bibnamefont {Breit}}\ and\ \bibinfo {author} {\bibfnamefont {I.~I.}\ \bibnamefont {Rabi}},\ }\bibfield  {title} {\bibinfo {title} {Measurement of {Nuclear} {Spin}},\ }\href {https://doi.org/10.1103/PhysRev.38.2082.2} {\bibfield  {journal} {\bibinfo  {journal} {Physical Review}\ }\textbf {\bibinfo {volume} {38}},\ \bibinfo {pages} {2082} (\bibinfo {year} {1931})}\BibitemShut {NoStop}%
\bibitem [{\citenamefont {Steck}(2003)}]{steck}%
  \BibitemOpen
  \bibfield  {author} {\bibinfo {author} {\bibfnamefont {D.}~\bibnamefont {Steck}},\ }\href@noop {} {\bibinfo {title} {Rubidium 87 {D} line data}} (\bibinfo {year} {2003})\BibitemShut {NoStop}%
\bibitem [{\citenamefont {Kay}(2013)}]{Kay3}%
  \BibitemOpen
  \bibfield  {author} {\bibinfo {author} {\bibfnamefont {S.~M.}\ \bibnamefont {Kay}},\ }\href@noop {} {\emph {\bibinfo {title} {Fundamentals of statistical signal processing}}},\ Prentice Hall signal processing series\ (\bibinfo  {publisher} {Prentice Hall},\ \bibinfo {address} {Englewood Cliffs, N.J},\ \bibinfo {year} {2013 - 2013})\BibitemShut {NoStop}%
\bibitem [{\citenamefont {Aitken}(1936)}]{Aitken2}%
  \BibitemOpen
  \bibfield  {author} {\bibinfo {author} {\bibfnamefont {A.~C.}\ \bibnamefont {Aitken}},\ }\bibfield  {title} {\bibinfo {title} {Iv.—on least squares and linear combination of observations},\ }\href {https://doi.org/10.1017/S0370164600014346} {\bibfield  {journal} {\bibinfo  {journal} {Proceedings of the Royal Society of Edinburgh}\ }\textbf {\bibinfo {volume} {55}},\ \bibinfo {pages} {42–48} (\bibinfo {year} {1936})}\BibitemShut {NoStop}%
\end{thebibliography}%

\end{document}